\documentclass{article}

\usepackage{arxiv}

\usepackage[utf8]{inputenc} 
\usepackage[T1]{fontenc}    
\usepackage{hyperref}       
\usepackage{url}            
\usepackage{booktabs}       
\usepackage{amsfonts}       
\usepackage{nicefrac}       
\usepackage{microtype}      
\usepackage{lipsum}		
\usepackage{subcaption} 
\usepackage{graphicx}
\usepackage{natbib}
\usepackage{doi}
\usepackage{amssymb}
\usepackage{amsmath}
\usepackage{amsthm}
\usepackage{mathtools}
\usepackage{bm}
\usepackage{dirtytalk}
\usepackage{verbatim}
\usepackage{bm}
\usepackage{parskip}
\usepackage{dsfont}
\usepackage{algorithm}
\usepackage{algpseudocode}
\usepackage[dvipsnames]{xcolor}

\usepackage{float}

\newcommand{\R}{\mathbb{R}}

\newcommand{\bk}{\bm{k}}

\newcommand{\bx}{\bm{x}}

\title{Dipole-lets: A new multiscale decomposition for MR Phase and Quantitative Susceptibility Mapping}
\author{ \href{https://orcid.org/0009-0005-4522-0343}{\includegraphics[scale=0.06]{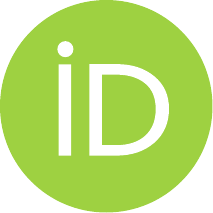}\hspace{1mm}Ignacio Contreras-Z\'u\~niga}\\
	Biomedical Imaging Center\\
	Pontificia Universidad Católica de Chile\\
	\texttt{iscontreras@uc.cl} \\
	\And
	 \href{https://orcid.org/0000-0002-2996-8141}{\includegraphics[scale=0.06]{orcid.pdf}\hspace{1mm}Mathias Lambert} \\
	Department of Electrical Engineering\\
	Biomedical Imaging Center\\
	Pontificia Universidad Católica de Chile\\
	\texttt{mglambert@uc.cl} \\
	\AND
	 \href{https://orcid.org/0000-0003-1671-5839}{\includegraphics[scale=0.06]{orcid.pdf}\hspace{1mm}Benjam\'in Palacios} \\
	Department of Mathematics\\
	Pontificia Universidad Católica de Chile \\
	\texttt{benjamin.palacios@uc.cl} \\
	\And
	Cristian Tejos \\
	Department of Electrical Engineering \\
    Biomedical Imaging Center\\
    Millennium Institute for Intelligent Healthcare Engineering\\
	Pontificia Universidad Católica de Chile \\
	\texttt{ctejos@uc.cl} \\
	\And
	 \href{https://orcid.org/0000-0002-1196-6703}{\includegraphics[scale=0.06]{orcid.pdf}\hspace{1mm}Carlos Milovic} \\
	Department of Electrical Engineering \\
    Biomedical Imaging Center\\
	Pontificia Universidad Católica de Chile \\
	\texttt{cmilovic@uc.cl} \\
}

\renewcommand{\shorttitle}{\textit{Dipole-lets} for MR Phase and QSM}

\hypersetup{
pdftitle={Dipole-lets for Quantitative Susceptibility Mapping},
pdfsubject={physics.med-ph, eess.SP},
pdfauthor={Contreras-Z\'u\~niga, Mathias Lambert, Benjam\'in Palacios, Cristian Tejos, Carlos Milovic},
pdfkeywords={Quantitative Susceptibily Mapping, Multiscale Analysis, Inverse Problems},
}

\begin{document}
\maketitle

\begin{abstract}
 Nondipolar phase contributions can generate severe streaking artifacts during quantitative susceptibility mapping (QSM) inversion. We propose Dipole-lets, a dipole-adapted multiscale decomposition designed to identify phase components associated with streaking artifacts before and during susceptibility inversion. Dipole-lets combine an undecimated radial decomposition with an angular partition based on proximity to the magic cone. The resulting coefficients emphasize phase components near the magic cone, where nondipolar contributions may become relatively prominent. This information was incorporated into QSM reconstruction through a data-driven fidelity weight and a Dipole-let-based regularizer that models an additional nondipolar phase component. The methods were evaluated using modified simulated QSM data and an in vivo dataset. The data-driven weighting reduced streaking artifacts while preserving anatomical detail and improved quantitative reconstruction metrics compared with magnitude-weighted TV. The Dipole-let regularizer achieved streaking suppression comparable to L1-QSM and lower reconstruction error in the modified simulated dataset. In vivo experiments further demonstrated the applicability of the proposed approach to data with strong susceptibility-induced phase perturbations. Dipole-lets provide a multiscale representation for characterizing phase components associated with streaking artifacts and incorporating this information into QSM reconstruction, reducing streaking while preserving relevant susceptibility structures, and providing a flexible basis for future QSM reconstruction methods.

\end{abstract}

%

\newpage

\section{Introduction}\label{sec:introduction}

In gradient-echo magnetic resonance imaging (GRE MRI), the phase evolves proportionally to magnetic field perturbations. Assuming these perturbations originate from tissue magnetization, Quantitative Susceptibility Mapping (QSM) estimates the spatial distribution of magnetic susceptibility by inverting a convolutional model in which susceptibility is represented as a distribution of non-interacting dipoles \cite{brown2014magnetic,qsm2024recommended,choi2014inverse}. This inverse problem is ill-posed because the dipole kernel contains a conical surface of zeros, known as the magic cone, which prevents exact inversion \cite{salomir2003fast,marques2005application}. Regularization or data-driven methods are therefore commonly used to obtain stable solutions \cite{kee2017quantitative}.

Early methods such as truncated k-space division (TKD) \cite{shmueli2009magnetic,wharton2010susceptibility} demonstrated the feasibility of in vivo susceptibility mapping but revealed severe streaking artifacts. These are non-local and arise from the amplification of frequencies near the dipole kernel zeroes during inversion. Their severity increases with noise, preprocessing errors (e.g., phase unwrapping, residual background fields) \cite{schweser2017illustrated,qsm2024recommended}, and physical model mismatches such as anisotropy or intra-voxel dephasing, not accounted for by the QSM forward model \cite{schweser2018quantitative,li2017susceptibility,milovic20202016}. These can produce localized errors or low-frequency shadowing artifacts, substantially degrading susceptibility estimation. Consequently, effective QSM reconstruction requires both accurate physical modeling and artifact-aware inversion strategies.

Many QSM reconstruction methods reduce streaking through convex optimization \cite{bilgic2015sparse}, minimizing a data fidelity term together with a convex regularizer that promotes desirable features in the susceptibility map. These methods range from Tikhonov regularization to total variation (TV)-based approaches such as MEDI \cite{liu2013nonlinear}, which incorporates structural priors \cite{kee2018coherence} to weight the TV regularizer \cite{rudin1992nonlinear} and its faster ADMM implementation, FANSI \cite{milovic2018fast}. TV-based methods achieved the best performance in the QSM Reconstruction Challenge 2.0 \cite{qsm2021qsm}. More advanced approaches improve the data fidelity term using nonlinear complex projections \cite{liu2013nonlinear,milovic2018fast,polak2020nonlinear} or employ L1-norm formulations to reject phase outliers \cite{milovic2022streaking}. Despite these advances, the regularization remains largely isotropic and therefore ignores the directional nature of streaking artifacts. Consequently, suppressing strong streaking often requires aggressive regularization, resulting in oversmoothing and loss of anatomical detail.

Several approaches have attempted to exploit the geometry of the dipole kernel more directly. Early work interpreted streaking artifacts as a propagation of singularities associated with the wave equation \cite{choi2014inverse,zhou2017dipole}, motivating later streaking-reduction methods based on pseudodifferential operators \cite{palacios2017reducing,hormander2007analysis}. More recently, multiscale directional decompositions have also been investigated for QSM \cite{starck2010sparse,huang2023robust,ahn2020quantitative}, including shearlets \cite{labate2005sparse}, however, these decompositions are not explicitly adapted to the geometry of the dipole kernel, limiting their ability to selectively identify streaking-inducing phase components.

In this work, we present a new multiscale decomposition for QSM, called the Dipole-lets decomposition, which is constructed from the geometry of the dipole kernel by combining angular scaling in the frequency domain with an undecimated wavelet transform \cite{starck2007undecimated} for radial scaling. Unlike existing directional decompositions, Dipole-lets are designed around the dipole operator itself, enabling nondipolar phase components to be identified through their multiscale energy distribution while providing targeted tools for streaking artifact reduction.

\section{Theory}\label{sec:theory}

This section introduces the proposed Dipole-lets decomposition and its theoretical foundations. Following the standard continuous formulation of QSM derived from Maxwell's equations, we developed the proposed decomposition in the same mathematical setting.

\subsection{Inverse Problem in QSM and Streaking Artifacts}\label{sec:theoryQSM}
In GRE MRI, magnetic field perturbations are encoded in the phase of the measured complex signal
\begin{equation}
    s_{\text{TE}}(\bx) = M(\bx)e^{-R_{2}^{*}(\bx)\text{TE}}e^{i\gamma B_{0}\psi_{D}(\bx) \text{TE}}, \quad \bx\in\R^{3}
\end{equation}
\noindent
where $B_0$ is the main magnetic field strength, $\gamma$ is the gyromagnetic ratio, TE is the echo time, and $M(\bx)$ is the signal magnitude of the GRE signal at $\text{TE}=0$, and $R_{2}^{*}(\bx)$ is the effective transverse relaxation rate. The dipolar phase is
\begin{equation}
\psi_{D}:=d*\chi
\end{equation}
\noindent
where $d$ denotes the magnetic dipole kernel convolved with the magnetic susceptibility distribution $\chi$.
In the Fourier domain, convolution becomes point-wise multiplication, yielding the dipole kernel
\begin{equation}
    D(\bk) =\left(\frac{1}{3}-\frac{k_{z}^{2}}{\|\bk\|_{2}^{2}}\right), \quad \bk\in \R^{3}\backslash\{\bm{0}\}.
\end{equation}
Since
\begin{equation}
\frac{k_z}{\|\bk\|_{2}} = \cos(\theta),
\end{equation}
\noindent
the dipole kernel depends only on $\theta$, the orientation of the frequency vector with respect to the $z$-axis. Its zero level-set defines the conical surface
\begin{equation}
\label{eqn:char}
\Sigma :=
\left\{
\bk\in\mathbb{R}^{3} :
k_x^2+k_y^2=2k_z^2
\right\},
\end{equation}
\noindent
known as the \emph{magic cone}, corresponding to an angle of approximately $54.7^\circ$ with respect to $B_0$.
Because the dipole kernel vanishes on $\Sigma$, QSM is an ill-posed inverse problem, and exact inversion is unstable. Noise and nondipolar phase components are amplified during dipole inversion, particularly near the magic cone \cite{zhou2017dipole}, producing the characteristic streaking artifacts.
Accordingly, we model the measured local phase as
\begin{equation}
\psi=\psi_D+\psi_N,
\end{equation}
\noindent
where $\psi_D$ denotes the phase component described by the scalar dipole convolution model, and $\psi_N$ collects phase contributions not represented by this model. For simplicity, we refer to $\psi_N$ hereafter as the \emph{nondipolar phase component}.

The nondipolar phase component may include contributions from phase-unwrapping errors, residual background fields, susceptibility anisotropy, microstructural effects, and other deviations from the scalar dipole model. Thermal noise is not included in $\psi_N$, as it originates from noise in the measured complex MRI signal and is instead addressed through the reconstruction formulation.
\subsection{The Dipole-lets Decomposition for MR Phase Analysis}\label{sec:theoryDecomposition}
The geometry of the dipole kernel suggests that an effective multiscale representation should distinguish not only spatial frequencies but also their proximity to the magic cone. Such a representation is naturally suited to distinguish components according to their consistency with the dipole geometry, providing a basis for identifying nondipolar sources responsible for streaking artifacts before susceptibility inversion.

Wavelets efficiently separate spatial scales but do not explicitly account for the geometry of the dipole kernel. Shearlets partially overcome this limitation by introducing directional selectivity through wedge-like partitions of the Fourier domain, providing sparse representations of anisotropic features. Shearlet-based QSM regularization \cite{stiegeler2020shearlet}, including combinations with TGV \cite{stiegeler2021shearlet}, has shown promising performance. However, because shearlet orientations are not defined relative to the dipole kernel, coefficients associated with streaking artifacts may overlap with those representing anatomical susceptibility, limiting their ability to isolate components that violate the dipole model.

These observations motivate the proposed Dipole-lets decomposition, a multiscale representation specifically designed around the geometry of the dipole kernel. The transform combines radial frequency decomposition with an angular partition based on proximity to the magic cone.

\subsubsection{Radial Decomposition}\label{sec:theoryRadial}

The radial component follows the multiscale construction of the undecimated wavelet (starlet) transform \cite{starck2007undecimated}, which is implemented directly in the Fourier domain. It is built from a cascade of Gaussian low-pass filters, where each radial band captures the spectral content removed at a given scale.
Define the Gaussian bandwidth by
\begin{equation}
\label{eqn:radialSigma}
\sigma_{j}:=\sigma 2^{j},
\qquad
\sigma>0,
\quad
j=0,\ldots,J-1,
\end{equation}
The corresponding Gaussian low-pass filter is
\begin{equation} 
\label{eqn:radialGaussian} G_j(\mathbf{k}) := \exp\left( -\frac{1}{2}\sigma_{j}^2 \|\mathbf{k}\|_2^2 \right). 
\end{equation}
The radial band-pass filters are constructed from the Gaussian low-pass filters as:
\begin{equation} 
\label{eqn:radialWindow} R_j(\mathbf{k}) := \left( \prod_{\ell=0}^{j-1} G_\ell(\mathbf{k}) \right) \left( 1-G_j(\mathbf{k}) \right), \quad j=0,\ldots,J-1, 
\end{equation}
\noindent
Since an empty product equals one,
\begin{equation}
R_0(\mathbf{k}) = 1-G_0(\mathbf{k})
\end{equation}
After the final scale, the remaining low-pass residual is
\begin{equation} 
\label{eqn:radialResidual} 
R_{\mathrm{res}}(\mathbf{k}) := \prod_{\ell=0}^{J-1} G_\ell(\mathbf{k}).
\end{equation}
The radial band-pass filters, together with the low-pass residual (which for further analysis we do not decompose angularly), form a smooth partition of unity:
\begin{equation} 
\label{eqn:radialPartition} 
\sum_{j=0}^{J-1} R_j(\mathbf{k}) + R_{\mathrm{res}}(\mathbf{k}) = 1.
\end{equation}
\subsubsection{Angular Decomposition}\label{sec:theoryAngular}
The angular component partitions the Fourier domain according to proximity to the magic cone, quantified by the absolute value of the dipole kernel $D(\mathbf{k})$. Let
\begin{equation}  
\delta_m:=\delta 2^{m}, \qquad \delta>0, \quad m=0,\ldots,M-2, 
\end{equation}
\noindent
and define the angular low-pass filter
\begin{equation}
\label{eqn:dipoleletAngularLowpass}
\begin{aligned}
    H_m(\mathbf{k}) := \frac{1}{2} \left( 1 - \operatorname{erf}\left( \frac{|D(\mathbf{k})| - \delta_{m}}{w} \right) \right), \\
    m=0,\ldots,M-2.
\end{aligned}
\end{equation}
\noindent
where $w$ controls the transition width. The error function was chosen instead of a Gaussian profile because it provides a sharper transition around the threshold $\delta_m$, while allowing the transition width to be controlled explicitly by $w$. The error function is defined as
\begin{equation}
\label{eqn:errorFunctionDef}
\operatorname{erf}(x) = \frac{2}{\sqrt{\pi}} \int_{0}^{x} e^{-y^2} \mathrm{d}y.
\end{equation}
This filter provides a smooth transition from $1$ to $0$ around the threshold
$\delta_m$, isolating frequencies close to the magic cone. Defining $H_{-1}(\mathbf{k}) := 0$ and $H_{M-1}(\mathbf{k}):=1$, the angular band-pass filters are
\begin{equation} \label{eqn:angularWindow} 
A_m(\mathbf{k}) := H_m(\mathbf{k}) - H_{m-1}(\mathbf{k}), \quad m=0,\ldots,M-1. 
\end{equation}
\noindent
Thus, the angular band-pass filters form a smooth partition of unity,
\begin{equation} \label{eqn:angularPartition} 
\sum_{m=0}^{M-1} A_m(\mathbf{k}) = H_{M-1}(\mathbf{k}) - H_{-1}(\mathbf{k}) = 1. 
\end{equation}
\subsubsection{The Dipole-lets Decomposition}\label{sec:theoryDipolelet}
\begin{figure*}[t]
    \centerline{
    \includegraphics[width=500pt] {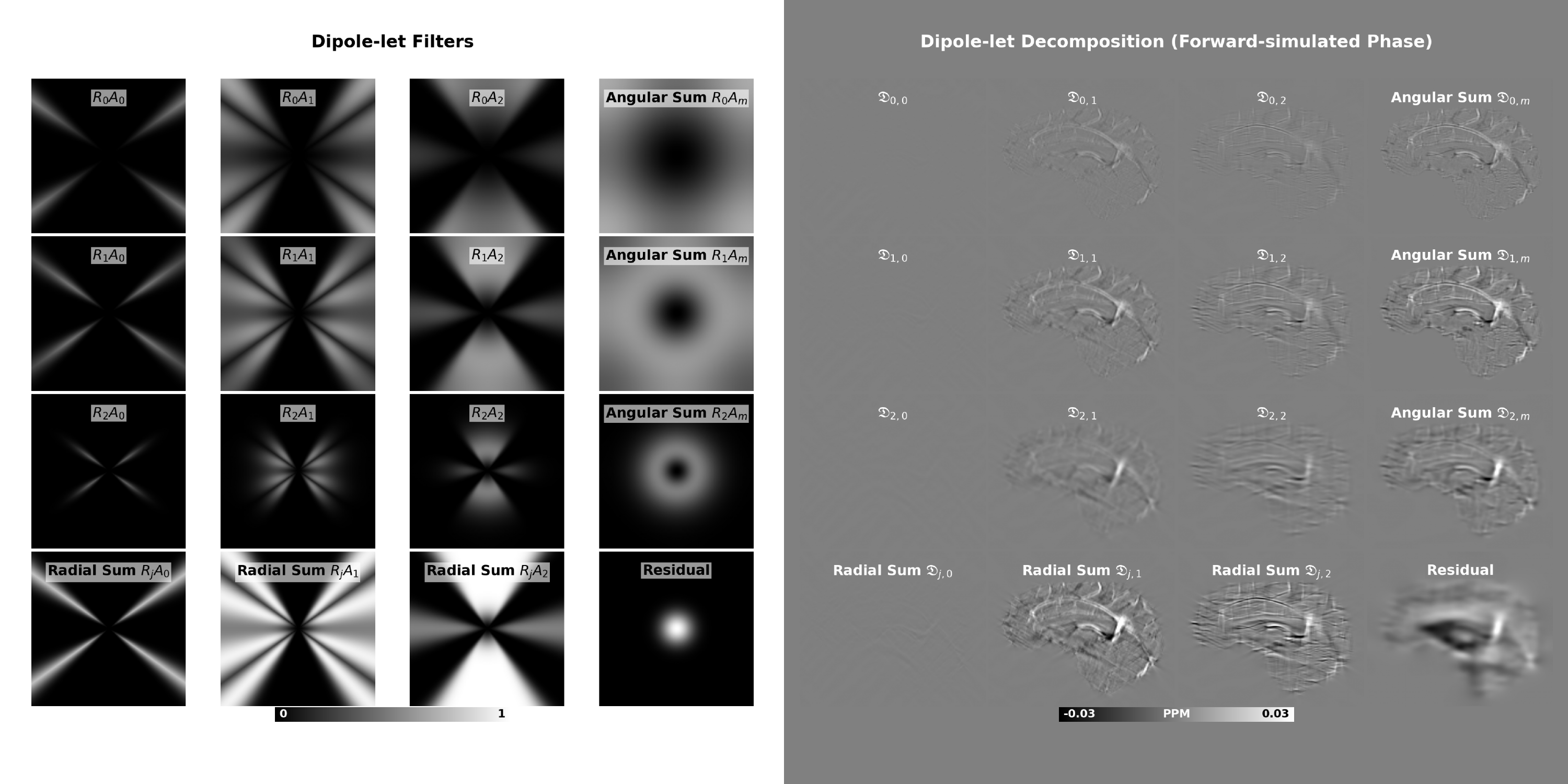}}
     \caption{Visualization of the Dipole-lets decomposition $\mathfrak{D}_{j,m}\psi$. $3\times3$ radial/angular scales, $\sigma=0.5$, $\delta = 0.05$, $w=0.5$. (Left) Dipole-let filters in frequency domain, Sagittal view. (Right) Dipole-lets decomposition of forward-simulated phase. As a forward-simulated phase is consistent with the dipole convolution, near-cone scales $\mathfrak{D}_{j,0}$, $j=0,1,2$, show negligible energy compared to others (see Right Figure, first column).}
    \label{fig:dipoleletDecomposition}
\end{figure*}

The Dipole-lets decomposition combines the radial and angular filter banks defined in the previous subsections to obtain a multiscale representation of an input image $\psi$. Let $\mathcal{F}$ and $\mathcal{F}^{-1}$ denote the forward and inverse Fourier transforms, respectively. For each $j$ and $m$ we define the Dipole-let operator $\mathfrak{D}_{j,m}$ as
\begin{equation}
\label{eqn:dipoleletWindows}
\begin{aligned}
    \mathfrak{D}_{j,m}\psi:= \mathcal{F}^{-1}R_{j}A_{m}\mathcal{F}\psi, \quad\; & j=0,\dots J-1, \\
    &m=0,\dots,M-1.
\end{aligned}
\end{equation}
so that $\mathfrak{D}_{j,m}\psi$ denotes the dipole-let coefficient of $\psi$ at scale $(j,m)$.
As the radial residual is not decomposed angularly, the dipolelet residual is defined as: 
\begin{equation}
\label{eqn:dipoleletResidual}
\mathfrak{D}_{\mathrm{res}}\psi
:=
\mathcal{F}^{-1}
(
R_{\mathrm{res}}
\mathcal{F}\psi
).
\end{equation}
Because the radial and angular filters form partitions of unity, the original phase image is reconstructed as
\begin{equation}
\label{eqn:dipoleletReconstruction}
\psi
=
\sum_{j=0}^{J-1}
\sum_{m=0}^{M-1}
\mathfrak{D}_{j,m}\psi
+ \mathfrak{D}_{\mathrm{res}}\psi.
\end{equation}
The resulting transform is undecimated and therefore yields a redundant, translation invariant, multiscale representation. Exact reconstruction follows directly from the partition-of-unity property of the radial and angular filter banks. Figure~\ref{fig:dipoleletDecomposition} illustrates the corresponding frequency-domain filters. 

\subsection{Dipole-lets and Separating Streaking Artifacts}\label{sec:theorySeparating}

\begin{figure}
\centering
    \setlength{\lineskip}{0pt}
    \setlength{\parskip}{0pt}
    \setlength{\baselineskip}{0pt}%
\includegraphics[width = 250pt]{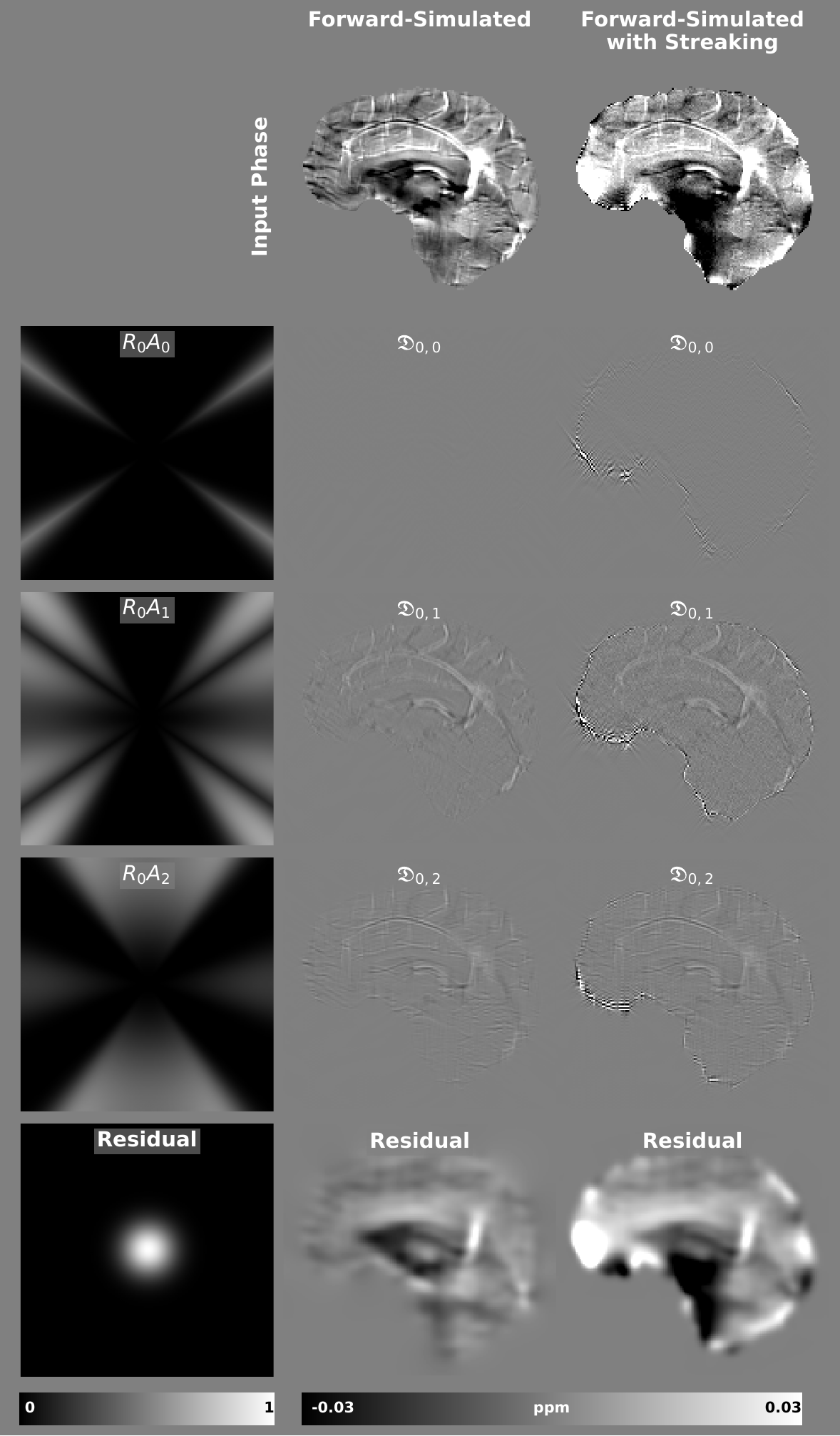}\\%
    \caption{Comparison of the first Dipole-let scale $\mathfrak{D}_{0,m}$, $m=0,1,2$ and residual $\mathfrak{D}_{\text{res}}$ for a purely dipolar phase (Forward-Simulated) image and the same image after adding a nondipolar phase component (Forward-Simulated with Streaking), including boundary errors due to unwrapping and background field removal remnants. While the dipolar image exhibits negligible energy in $\mathfrak{D}_{0,0}$, the added nondipolar component produces a strong response, and the multiscale nature of the decomposition allows to separate smooth background remnant in the residual from other coefficients, illustrating how the proposed decomposition separates nondipolar sources prior to dipole inversion, both high and low frequency ones.}
    \label{fig:WF}
\end{figure}

Here we provide the intuition underlying its ability to isolate nondipolar phase components prior to dipole inversion. Because the dipole kernel vanishes on the magic cone, purely dipolar fields are increasingly attenuated as frequencies approach the magic cone. Consequently, within the sufficiently narrow angular region defined by $A_0$, nondipolar components can become relatively more prominent than the dipolar contribution. For example, consider a single-voxel phase jump, which represents a localized nondipolar component. In the Fourier domain, its spectrum spreads across all scales with approximately constant magnitude. Consequently, the Dipole-let coefficients associated with the magic cone can exhibit significant energy. Since such nondipolar components are inconsistent with the dipole model, they can generate streaking artifacts during susceptibility inversion. This provides a mechanism for identifying potential sources of streaking artifacts, as illustrated in (Figure~\ref{fig:WF}).

\section{Methods}\label{sec:methods}

\subsection{Dipole-lets for Data-driven Fidelity Weights} \label{sec:methodsWeight}

Dipole-lets provide a voxel-wise energy map by accumulating the energy of coefficients from selected Dipole-let scales $\Omega$

\begin{equation}
\label{eqn:dipoleletEnergy}
\begin{aligned}
    E_{\Omega}(\bx):=
    \sum_{(j,m)\in\Omega}
    \left|\mathfrak{D}_{j,m}\psi(\bx)\right|^2, \quad
    \Omega \subseteq
    \{0,\ldots,J-1\}\times\{0,\ldots,M-1\}.
\end{aligned}
\end{equation}

\noindent
To isolate nondipolar components, $\Omega$ is limited to scales near $\Sigma$ (e.g., $m=0,1$). However, as nondipolar phase components are commonly sharp spatial discontinuities (phase jumps, unwrapping errors), their spectrum is broadband. Therefore, $\Omega$ should include multiple radial scales to capture the broadband spectral content of spatially localized nondipolar components.

We define a normalized and thresholded energy map
$\widehat{E}_{\Omega}$ by normalizing $E_{\Omega}$ by its maximum,
discarding values below a threshold $\tau$, and applying a fractional
exponent $0<\gamma<1$ to the retained values:

\begin{equation}
    \widehat{E}_{\Omega}(\bx) :=
    \begin{cases}
        \left(\frac{E_{\Omega}(\bx)}{\max\limits_{\bx}E_{\Omega}(\bx)}\right)^{\gamma}, & \text{if } \frac{E_{\Omega}(\bx)}{\max\limits_{\bx}E_{\Omega}(\bx)} > \tau \\
        0, & \text{otherwise.}
    \end{cases}
\end{equation}

Let $\mathbf{\widehat{E}_{\Omega}}$ be the discretized energy map. We incorporate this into magnitude weighting
\begin{equation}
\label{magnitudeWeight}
    \mathbf{M}^{\star} = \frac{\sum_{i=1}^{N}{\mathbf{M}_{i}^{2}}\cdot\text{TE}_{i}}{\sum_{i=1}^{N}{\mathbf{M}_{i}}\cdot\text{TE}_{i}}
\end{equation}
where $\mathbf{M}_{i}$ and TE are the magnitude and echo-time of the $i$-th echo, to create the data-driven fidelity weight map as
\begin{equation}
\label{eqn:DipoleletsWeight}
    \mathbf{W}_{\mathfrak{D}}(\bx):= \mathbf{M}^{\star}(\mathbf{1}- \mathbf{\widehat{E}_{\Omega}})
\end{equation}
which can then be used in a weighted TV solver
\begin{equation}
    \label{weightedTV}
            \begin{aligned}
                \underset{\bm{\chi}}{\text{minimize}} \quad & \frac{1}{2}\|\mathbf{W}_{\mathfrak{D}}(\mathbf{F}^{\mathsf{H}}\mathbf{D}\mathbf{F}\bm{\chi}-\bm{\psi})\|_{2}^{2}+\alpha \|\bm{\chi}\|_{\text{TV}}, \alpha>0
            \end{aligned}
\end{equation}
\noindent
where $\|\cdot\|_{\text{TV}}$ denotes 
the total variation norm. Thus, we create a magnitude-weight that is robust to the nondipolar phase components, using only the data information provided by the Dipole-let decomposition. See the Appendix for additional details.
\subsection{Dipole-lets as Regularizer}
\label{sec:methodsRegularizer}
Streaking artifacts arise from nondipolar phase components that are not adequately modeled by the dipole inversion. We model these components through the nondipolar phase term $\bm{\psi}_{N}$ and introduce an additional Dipole-let-based regularization term that promotes dipole-inconsistent structure to be absorbed in $\bm{\psi}_N$ 
\begin{equation}
\label{eqn:dipoleletRegularizer}
    \begin{aligned}
                \underset{\bm{\chi},\bm{\psi}_{N}}{\text{minimize}} \quad &\frac{1}{2}\|\mathbf{M}(\mathbf{F}^{\mathsf{H}}\mathbf{D}\mathbf{F}\bm{\chi}+ \bm{\psi}_{N} - \bm{\psi})\|_{2}^{2} +
                \alpha \|\bm{\chi}\|_{\text{TV}}+\lambda\sum_{j=0}^{J-1}\sum_{m=0}^{M-1} \|\bm{\mathfrak{D}}_{j,m}\bm{\psi}_{N}\|_{1} + \lambda_{\text{res}}\|\bm{\mathfrak{D}}_{\text{res}}\bm{\psi}_{N}\|_{1}
            \end{aligned}
\end{equation}

\noindent
with $\alpha>0$, $\lambda>0$, and $\lambda_{\text{res}}>0$. We chose $J=1,M=2$, therefore defining the scales $(0,0), (0,1)$. A heuristic for choosing these parameters is $\lambda = 10\alpha$, $\lambda_{\text{res}}=15\alpha$, producing results close to optimal (see Section \ref{sec:methodsSIM2}). $\mathbf{M}$ is a binary ROI mask/magnitude-based weight

\subsection{Experimental Design}
\label{sec:methodsSIM2}

We designed two experiments to evaluate the data-driven fidelity weight and the Dipole-let regularizer, respectively.\\

\textbf{1. Dipole-lets as Data-Driven Fidelity Weights}: We evaluated whether the fidelity weight (Equation~\eqref{eqn:DipoleletsWeight}) identifies regions with strong near-cone energy, where nondipolar phase contributions may become relatively prominent, and to reduce the streaking generated by these contributions during dipole inversion. We used the SIM2SNR1 dataset from the 2019 QSM Reconstruction Challenge (RC2) \cite{marques2005application,qsm2021qsm}. The supplied field map contains a simulated calcification, corresponding to a strong negative susceptibility source that generates a large local field perturbation. Furthermore, we added a spatially localized parallelepiped phase artifact with an amplitude of $1$ Hz and sharp boundaries near the pons to simulate an extreme nondipolar phase error.

Weighted-TV reconstructions using the proposed Dipole-let-based fidelity weight (Equation~\eqref{eqn:DipoleletsWeight}) were compared with weighted-TV reconstructions using conventional magnitude weighting (Equation~\eqref{magnitudeWeight}). Both weighting strategies were evaluated using the original SIM2SNR1 field map and the modified field map containing the added artifact. 

\textbf{2. Dipole-lets as Regularizer}: This comparison was designed to asses whether the proposed regularizer (Equation~\eqref{eqn:dipoleletRegularizer}) reduces streaking induced by nondipolar phase components while preserving anatomical information, compared with state-of-the-art QSM reconstruction methods. Using the same modified SIM2SNR1 dataset containing the added nondipolar phase artifact described in the previous experiment, reconstructions obtained with the Dipole-let regularizer were compared with magnitude-weighted TV \cite{milovic2018fast}, L1-QSM \cite{milovic2022streaking}, and the deep learning methods xQSM \cite{gao2021xqsm} and QSMnet+ \cite{jung2020exploring}. Because the ground-truth susceptibility map is available for this dataset, regularization parameters for the iterative methods (TV, L1-QSM, and Dipole-lets) were selected to minimize RMSE. We additionally compared the RMSE-optimized Dipole-let reconstruction with that obtained using the heuristic parameter-selection strategy described in Section~\ref{sec:methodsRegularizer} (see Appendix).

We additionally evaluated the method in vivo using data acquired from a patient with cerebral hemorrhage on a 3T Siemens Magnetom Trio Tim scanner (Siemens Healthcare, Erlangen, Germany), using a 12-channel phased-array head coil and a six-echo GRE sequence. Acquisition parameters included: $\text{TE}1/\Delta\text{TE}/\text{TR} = 4.92/4.92/35 \text{ms}$, flip angle = $15^\circ$, matrix size = $232 \times 288 \times 64$, voxel dimensions = $0.8 \times 0.8 \times 2\text{ mm}^3$, and $T\text{acq} = 4{:}51$ min. Data preprocessing involved nonlinear multi-echo phase combination \cite{liu2013nonlinear}, followed by background field removal using PDF \cite{liu2011novel}. Residual background fields were further reduced using the harmonic phase estimation approach of WH-QSM \cite{milovic2019weak} prior to dipole inversion. As no ground truth is available for this dataset, optimization parameters for the iterative methods were selected using an L-curve followed by visual tuning. Data were transformed to isotropic resolution for inference with the deep learning methods.

For both experiments, the proposed optimization-based methods were implemented using ADMM \cite{boyd2011distributed,milovic2018fast}, based on the FANSI solver \cite{milovic2018fast} (see Appendix). Reconstruction quality was evaluated using RMSE and XSIM \cite{milovic2025xsim}. Data acquisition was approved by the local institutional review board.

\section{Results}\label{sec:results}
\subsection{Dipole-lets as Data-Driven Fidelity Weights}\label{sec:resultsWeight}
\begin{figure*}
\centerline{
 \includegraphics[width=342pt]{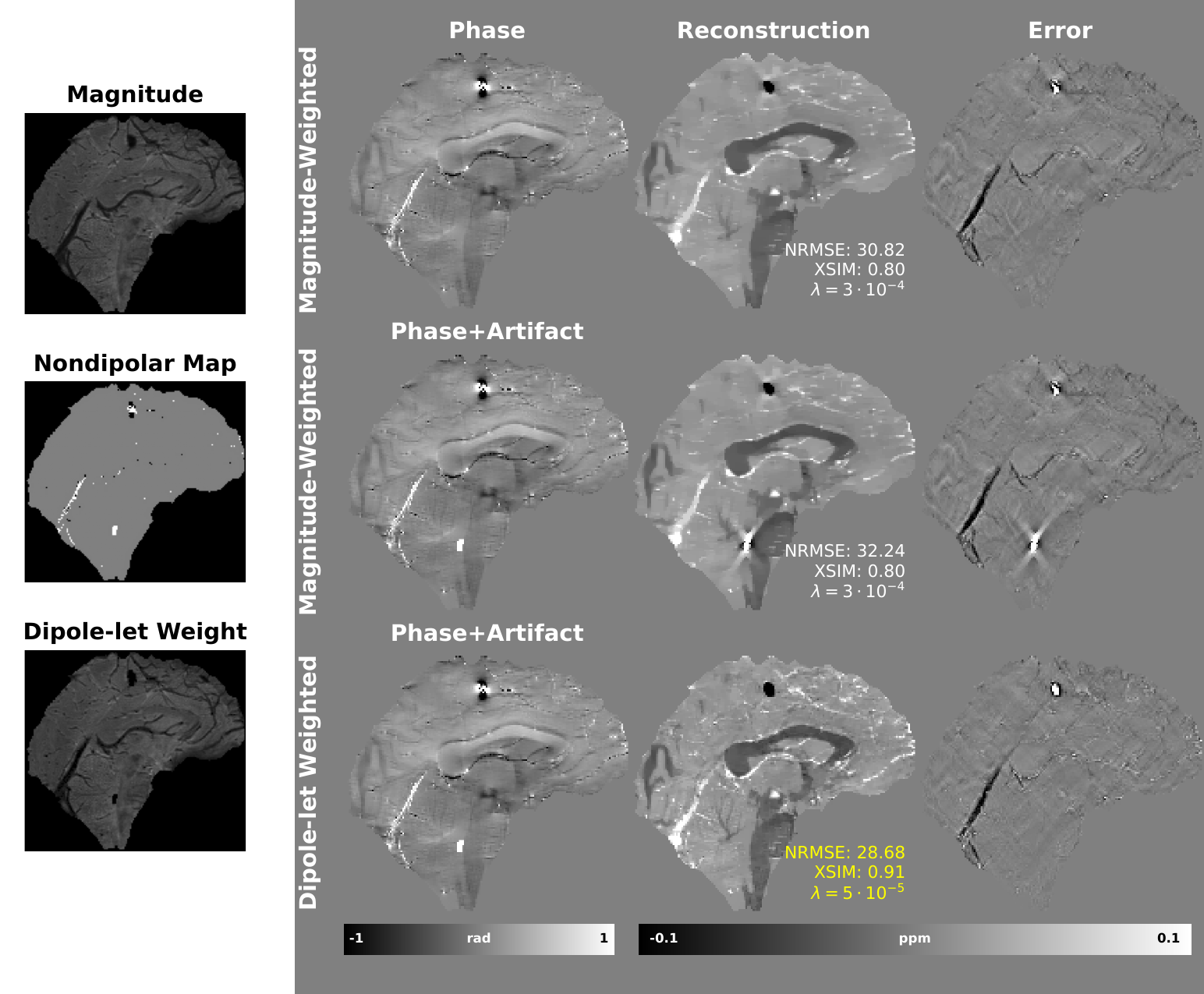}}
 \caption{Reconstruction for the SIM2SNR1 data with added artifact and the effect of the weight: Linear TV with magnitude-weighted and Dipole-let-weighted alongside the estimated nondipolar map to create the data-drive weight are shown. The Dipole-lets decomposition $\mathfrak{D}_{j,m}\psi$ is a $5\times3$ radial/angular decomposition, and we choose scale $\{(j,m)\}=\{(0,0),(0,1),(1,0),(1,1),(2,0),(2,1)\}$ to generate the data-driven weight, and all algorithms were stopped at 1500 iterations. Estimate of the high-frequency nondipolar phase contribution is shown (Nondipolar Map), computed by subtracting the Dipole-let residual from the measured phase, and then multiplying by the spatial energy: $\mathbf{\widehat{E}_{\Omega}} (\bm{\psi} - \bm{\mathfrak{D}}_{\text{res}}\bm{\psi})$.
 Note that both the main calcification (top of the brain) and the added artifact streaking are suppressed by the effect of the weight. See Appendix for the Dipole-let decomposition of the data.}
 \label{fig:weight}
\end{figure*}

The proposed data-driven weight improved streaking suppression compared with magnitude-weighted TV (Figure~\ref{fig:weight}). Importantly, this improvement was not accompanied by indiscriminate smoothing. While magnitude-weighted TV reduced streaking artifacts, it also resulted in loss of anatomical detail, including the great cerebral vein (vein of Galen). By contrast, the Dipole-let-based weighting improved the quantitative reconstruction scores, reducing NRMSE and increasing XSIM, while preserving fine anatomical detail. The selected Dipole-let scales identified regions with high near-cone energy associated with both the added phase artifact and the calcification, enabling streaking attenuation without substantially altering the overall calcification morphology.

\subsection{Dipole-lets as Regularizer}
\label{sec:resultsRegularizer}
\begin{figure*}
\centerline{\includegraphics[width=500pt]{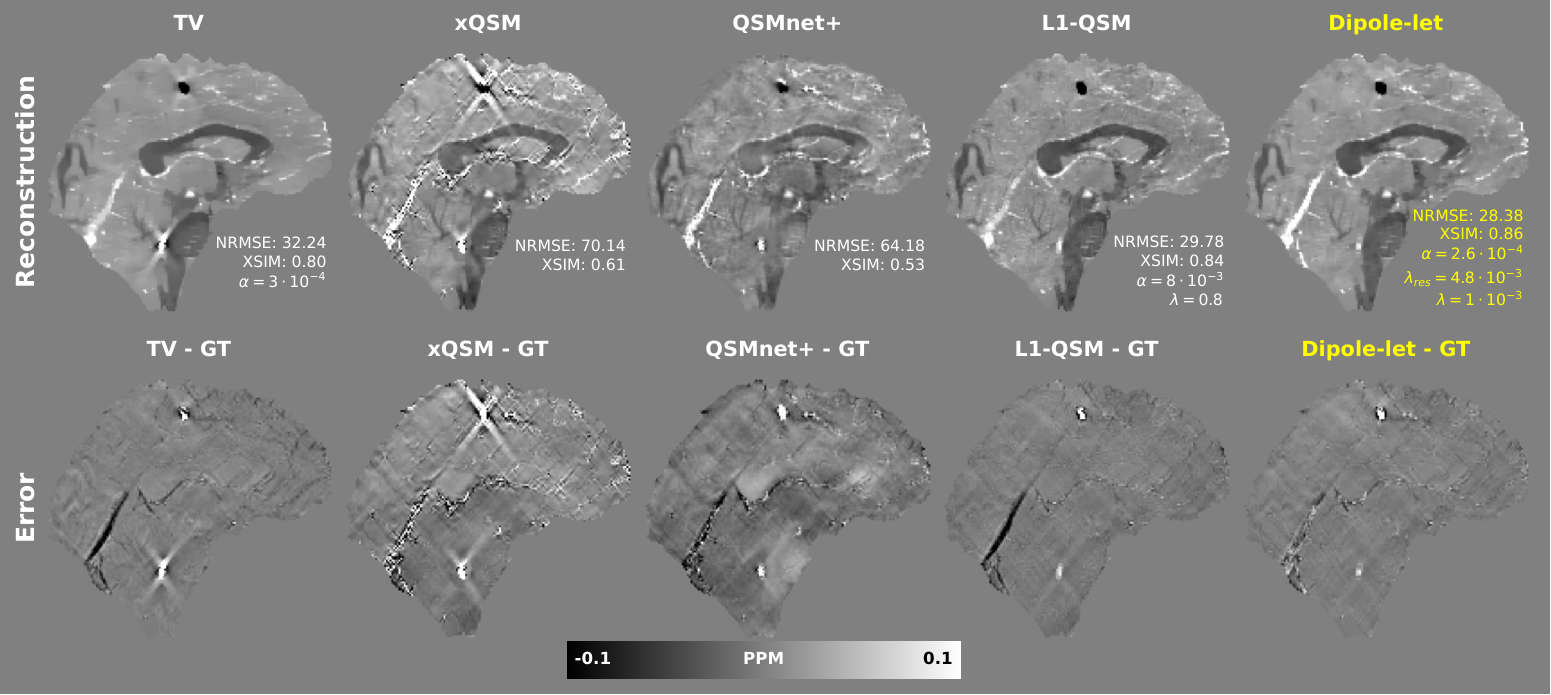}}
    \caption{Reconstruction for the SIM2SNR1 data with added artifact for different reconstruction methods, TV/L1-QSM are magnitude-weighted, Dipole-lets are ROI weighted. Dipole-lets use a $1\times 2$ angular/radial decomposition. Iterative methods are reported at best results. See Appendix for RMSE, XSIM and convergence evolution.}
   \label{fig:regularizerRC2}
\end{figure*}

The modified SIM2SNR1 dataset introduced an additional challenge for the evaluated reconstruction methods. As shown in Figure~\ref{fig:regularizerRC2}, xQSM reconstructions were strongly affected by streaking artifacts, whereas QSMnet+ reduced streaking at the cost of altering the contrast of surrounding tissues. Magnitude-weighted TV improved the reconstruction metrics, reducing NRMSE and increasing XSIM compared with both xQSM and QSMnet+. However, its streaking suppression was accompanied by a loss of relevant anatomical detail. Dipole-lets and L1-QSM achieved the best quantitative reconstruction performance, with comparable suppression of streaking artifacts. Dipole-lets yielded a lower reconstruction error than L1-QSM.

\begin{figure*}
\centerline{\includegraphics[width=500pt]{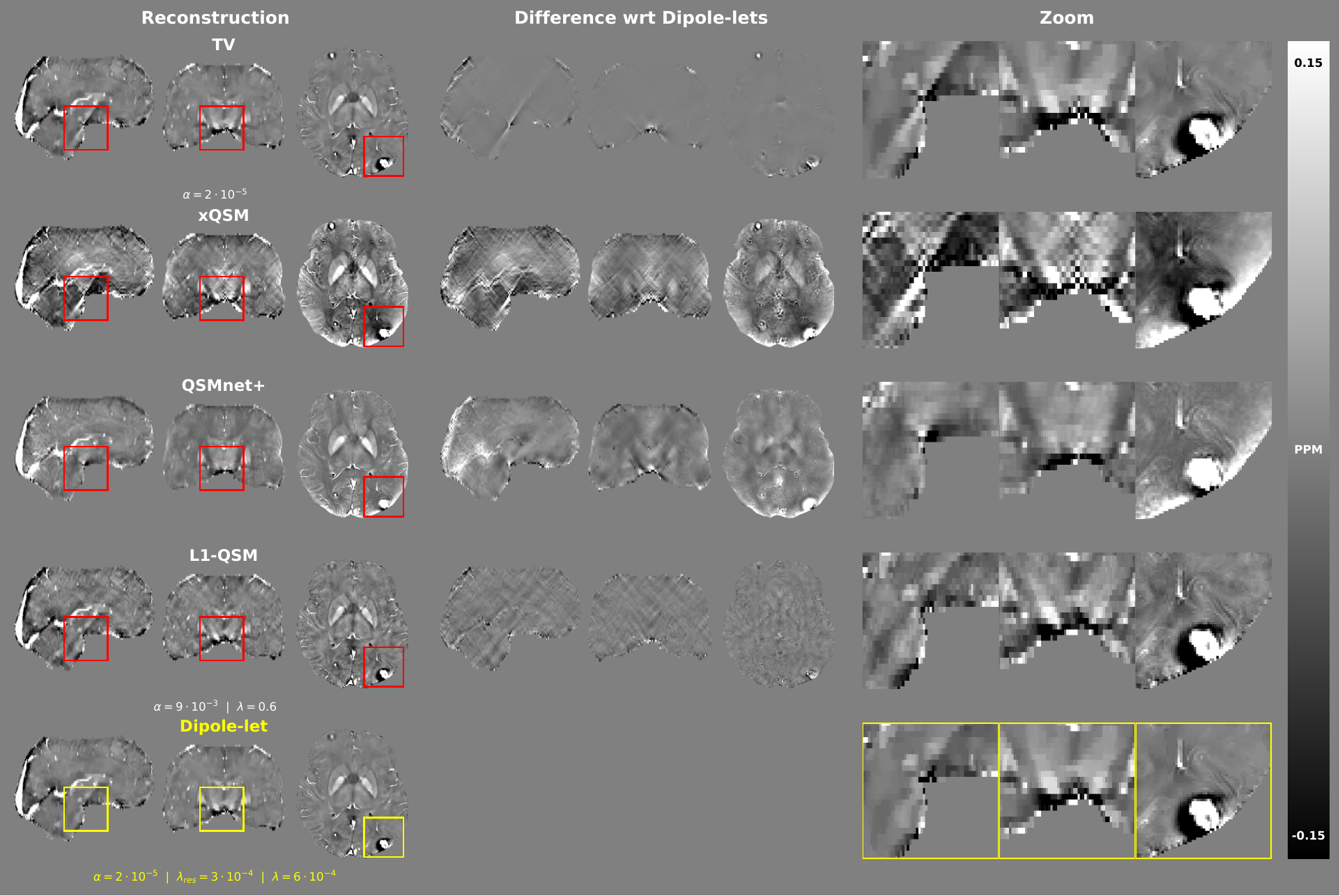}}
    \caption{Reconstruction for the in vivo phase for different reconstruction methods, TV/L1-QSM are magnitude-weighted, Dipole-lets is ROI weighted. Iterative methods were stopped at $500$ iterations.}
   \label{fig:regularizerRC1}
\end{figure*}

For the in vivo hemorrhage dataset (Figure~\ref{fig:regularizerRC1}), TV, L1-QSM, and Dipole-lets showed visually comparable streaking suppression. Among these methods, Dipole-lets showed better preservation of tissue contrast, while maintaining a similar apparent noise level.

\section{Discussion}
\label{sec:discussion}
Streaking suppression remains one of the main challenges in QSM, particularly when phase data contain components that are not well described by the scalar dipole model. These include pre-processing errors, imperfect background field removal, susceptibility anisotropy, and other sources of model mismatch. Such nondipolar contributions can generate streaking artifacts through the ill-posed inversion, which are difficult to suppress without degrading anatomical detail.

In this work, we introduced a multiscale dipole-adapted decomposition designed to analyze the local phase before and during susceptibility inversion. The Dipole-let transform provides a means to identify phase components potentially associated with model mismatch by analyzing their energy distribution across radial scales and their proximity to the magic cone, without explicitly modeling the underlying nondipolar source. This information was used to construct a data-driven fidelity weight and a regularizer, both of which reduced streaking while preserving susceptibility structures. The Dipole-let regularizer achieved performance comparable to the streaking-robust L1-QSM method, with lower reconstruction error in the modified SIM2SNR1 experiment.

One possible explanation for this behavior is that the Dipole-let representation provides information about artifact-prone phase components according to their proximity to the zeros of the dipole kernel. This allows the regularizer to retain an $\ell_{2}$ data-fidelity term while explicitly modeling an additional nondipolar phase component. However, the present experiments do not isolate the respective contributions of the Dipole-let representation and the choice of data-fidelity norm. The proposed decomposition could also be extended to nonlinear data-fidelity formulations operating on the complex MRI signal, where Gaussian noise assumptions are more appropriate.

The data-driven weights depend on the selection of radial and angular scales, which were visually tuned in this study. Similarly, Dipole-let regularization introduces additional optimization parameters, including the regularization weights and ADMM variables. The present validation is also limited to a modified simulated dataset and a single in vivo hemorrhage case; evaluation across larger and more diverse in vivo datasets will be necessary to establish the robustness and generalizability of the proposed methods.

\section{Conclusions}
\label{sec:conclusion}

We introduced Dipole-lets, a dipole-adapted multiscale decomposition designed to identify streaking-inducing phase components according to their radial scale and proximity to the magic cone. The proposed representation provides a flexible decomposition for incorporating information about potential nondipolar phase contributions into QSM reconstruction without explicitly modeling the underlying sources of model mismatch. The proposed data-driven weighting strategy reduced streaking artifacts during dipole inversion while preserving relevant susceptibility structures.

As a regularizer, Dipole-lets provide a competitive alternative to streaking-robust approaches such as L1-QSM, showing comparable streaking suppression and lower reconstruction error in the modified SIM2SNR1 experiment while retaining an $\ell_{2}$ data-fidelity formulation. Beyond optimization-based reconstruction, Dipole-lets may also support future deep learning pipelines by identifying artifact-prone phase components and providing a task-specific multiscale representation for improving robustness to nondipolar errors and assessing QSM reconstruction quality.

\section{Acknowledgments}\label{sec:acknowledgments}
We thank ANID Fondecyt de Iniciacion 11240279, Fondecyt 1231535, Fondecyt 1251207, and the Millennium Institute for Intelligent Healthcare Engineering (ICN2021\_004).








\bibliographystyle{unsrt}
\bibliography{references}
\newpage
\section{Appendix}
\subsection{Algorithm Description}\label{sec:algorithmDescription}
Here we explain the Dipole-let Regularizer. Our region of interest is discretized in a grid of size $N_{x}\times N_{y}\times N_{z}$ and let the number of voxels be $n = N_{x}N_{y}N_{z}$. The measured phase is $\bm{\psi}=\mathbf{F}^{\mathsf{H}}\mathbf{D}\mathbf{F}\bm{\chi}+\bm{\psi}_{N}$, where $\bm{\chi},\bm{\psi},\bm{\psi}_{N}\in \mathbb{R}^{n}$ are susceptibility, measured phase, and nondipolar phase vectors, $\mathbf{D}\in \mathbb{R}^{n\times n}$ is the diagonal dipole matrix in frequency and $\mathbf{F},\mathbf{F}^{\mathsf{H}} \in\mathbb{R}^{n\times n}$ are forward and conjugate transpose DFT matrices, and $\mathbf{M}\in \mathbb{R}^{n\times n}$ is a diagonal ROI magnitude-weighting matrix. For the following objective function:
\begin{equation}
\label{eqn:dipoleletRegularizerAp}
\begin{aligned}
\underset{\bm{\chi},\bm{\psi}_{N}}{\text{minimize}} \quad & \frac{1}{2}\|\mathbf{M}(\mathbf{F}^{\mathsf{H}}\mathbf{D}\mathbf{F}\bm{\chi}+ \bm{\psi}_{N} - \bm{\psi})\|_{2}^{2} + \alpha \|\bm{\chi}\|_{\text{TV}} + \lambda\sum_{j=0}^{J-1}\sum_{m=0}^{M-1} \|\bm{\mathfrak{D}}_{j,m}\bm{\psi}_{N}\|_{1} + \lambda_{\text{res}}\|\bm{\mathfrak{D}}_{\text{res}}\bm{\psi}_{N}\|_{1}
\end{aligned}
\end{equation}
the corresponding Augmented Lagrangian for the Dipole-let Regularizer is:
\begin{equation}
\label{eqn:dipoleletLagrangian}
\begin{aligned}
\mathcal{L}(\bm{\chi},\bm{\psi}_{N},\mathbf{z}_{1},\mathbf{z}_{2},\mathbf{z}_{\text{dip}}, \mathbf{s}_{1}, \mathbf{s}_{2},\mathbf{s}_{\text{dip}}) :=& \frac{1}{2} \|\mathbf{M} (\mathbf{z}_{2} - \mathbf{z})\|_2^2 + \alpha \|\mathbf{z}_1\|_1 \\
+& \frac{\mu_1}{2} \|\nabla \bm{\chi} - \mathbf{z}_1 + \mathbf{s}_1\|_2^2 + \frac{\mu_2}{2} \|\mathbf{F}^\mathsf{H} \mathbf{D} \mathbf{F} \bm{\chi} \\
+& \bm{\psi}_{N} - \mathbf{z}_{2} + \mathbf{s}_{2}\|_2^2 + \frac{\mu_{\text{dip}}}{2} \|\bm{\psi}_{N} - \mathbf{z}_{\text{dip}} + \mathbf{s}_{\text{dip}}\|_2^2\\
+& \lambda\sum_{j=0}^{J-1}\sum_{m=0}^{M-1} \|\bm{\mathfrak{D}}_{j,m}\mathbf{z}_{\text{dip}}\|_1 + \lambda_{\text{res}} \|\bm{\mathfrak{D}}_{\text{res}}\mathbf{z}_{\text{dip}}\|_1
\end{aligned}
\end{equation}
where $\mathbf{z}=\bm{\psi}$. The derivatives are computed in frequency by introducing the operator $\mathbf{E}=(\mathbf{E}_{1},\mathbf{E}_{2},\mathbf{E}_{3})$, where $\mathbf{E}_{d}$ is a diagonal discrete finite difference matrices for $d=(1,2,3)$. The k-th entry of each matrix is given by $\mathbf{E}_{d}({\mathbf{k}_{d}})=(1-e^{2\pi i \mathbf{k}_{d}/N_{d}})/\mathbf{v}_{d}$ where $N_d$ is the grid size and $\mathbf{v}_{d}$ is the voxel size at dimension $d$, so that $\mathbf{E}^{\mathsf{H}}\mathbf{E} = \mathbf{E}_{1}^{\mathsf{H}}\mathbf{E}_{1}+\mathbf{E}_{2}^{\mathsf{H}}\mathbf{E}_{2}+\mathbf{E}_{3}^{\mathsf{H}}\mathbf{E}_{3}$. With this definition the anisotropic total variation regularizer is computed as $\|\bm{\chi}\|_{\text{TV}}= \|\nabla\bm{\chi}\|_{1}$ where $\nabla\bm{\chi}= \mathbf{F}^{\mathsf{H}}\mathbf{E}\mathbf{F}\bm{\chi}$. We present the ADMM algorithm for solving Equation \ref{eqn:dipoleletLagrangian}, based on the FANSI implementation. We use analogous notation:
 
\begin{algorithm}[htbp]
\caption{Dipole-let Regularizer}
\label{alg:dip}
\begin{algorithmic}[1]
\Require Phase $\mathbf{z}$, weight $\mathbf{M}$, dipole $\mathbf{D}$, TV param. $\alpha$, dipolelet param. $\lambda$, residual dipolelet param. $\lambda_{\text{res}}$, variables $\mu_1, \mu_2, \mu_{\text{dip}}$, tolerance $\epsilon$, $\mathbf{W}=\mathbf{M}^{2}$, max. iterations $N_{iter}$
\State \textbf{Set:} $\bm{\chi} = \mathbf{0}, \bm{\psi}_{N} = \mathbf{0}$, $\mathbf{z}_1 = \mathbf{0}, \mathbf{s}_1 = \mathbf{0}$, $\mathbf{z}_{\text{dip}} = \mathbf{0},  \mathbf{s}_{\text{dip}} = \mathbf{0}$, $\mathbf{s}_{2} = \mathbf{0}$, $\mathbf{z}_{2} = \frac{\mathbf{W} \mathbf{z}}{\mathbf{W} + \mu_2}$

\For{$t = 1$ \textbf{to} $N_{iter}$}
    \State $\bm{\chi}_{0} \gets \bm{\chi}$

    \State $\bm{\chi} \gets \mathbf{F}^\mathsf{H} \left( \frac{\mu_1 \mathbf{E}^\mathsf{H} \mathbf{F} (\mathbf{z}_1 - \mathbf{s}_1) + \mu_2 \mathbf{D}^\mathsf{H} \mathbf{F} (\mathbf{z}_{2} - \bm{\psi}_{N} - \mathbf{s}_{2})}{\mu_2 \mathbf{D}^\mathsf{H} \mathbf{D} + \mu_1 \mathbf{E}^\mathsf{H} \mathbf{E}} \right)$

    \State $\bm{\psi}_{N} \gets \frac{\mu_2 (\mathbf{z}_{2} - \mathbf{F}^\mathsf{H} \mathbf{D} \mathbf{F} \bm{\chi} - \mathbf{s}_{2}) + \mu_{\text{dip}} (\mathbf{z}_{\text{dip}} - \mathbf{s}_{\text{dip}})}{\mu_2 + \mu_{\text{dip}}}$

    \State $\mathbf{v} \gets \bm{\psi}_{N} + \mathbf{s}_{\text{dip}}$
    \For{$j = 0$ \textbf{to} $J-1$, $m = 0$ \textbf{to} $M-1$}
        \State $\mathbf{c}_{j,m} \gets \bm{\mathfrak{D}}_{j,m}\mathbf{v}$ 
        \State $\mathbf{c}_{j,m}^{\star} \gets \max\left(|\mathbf{c}_{j,m}| - \frac{\lambda}{\mu_{\text{dip}}}, \mathbf{0}\right) \text{sign}(\mathbf{c}_{j,m})$ 
    \EndFor
    \State $\mathbf{c}_{\text{res}} \gets \bm{\mathfrak{D}}_{\text{res}}\mathbf{v}$ 
    
    \State $\mathbf{c}_{\text{res}}^{\star} \gets \max\left(|\mathbf{c}_{\text{res}}| - \frac{\lambda_{\text{res}}}{\mu_{\text{dip}}}, \mathbf{0}\right) \text{sign}(\mathbf{c}_{\text{res}})$ 
    
    \State $\mathbf{z}_{\text{dip}} \gets \left(\mathbf{c}_{j,m}^{\star}+ \mathbf{c}_{\text{res}}^{\star}\right)$  

    \State $\text{Update} \gets \frac{100 \times \| \bm{\chi} - \bm{\chi}_{0} \|_2}{\| \bm{\chi} \|_2}$
    \If{$\text{Update} < \epsilon$}
        \State \textbf{break} 
    \EndIf

    \State $\mathbf{z}_1 \gets \max\left(|\nabla \bm{\chi} + \mathbf{s}_1| - \frac{\alpha}{\mu_1}, \mathbf{0}\right) \text{sign}(\nabla \bm{\chi} + \mathbf{s}_1)$

    \State $\mathbf{z}_{2} \gets \frac{\mathbf{W} \mathbf{z} + \mu_2 (\mathbf{F}^\mathsf{H} \mathbf{D} \mathbf{F} \bm{\chi} + \bm{\psi}_{N} + \mathbf{s}_{2})}{\mathbf{W} + \mu_2}$

    \State $\mathbf{s}_1 \gets \mathbf{s}_1 + \nabla \bm{\chi} - \mathbf{z}_1$
    \State $\mathbf{s}_{\text{dip}} \gets \mathbf{s}_{\text{dip}} + \bm{\psi}_{N} - \mathbf{z}_{\text{dip}}$
    \State $\mathbf{s}_{2} \gets \mathbf{s}_{2} + \mathbf{F}^\mathsf{H} \mathbf{D} \mathbf{F} \bm{\chi} + \bm{\psi}_{N} - \mathbf{z}_{2}$
\EndFor
\Ensure Estimated $\bm{\chi}^{\star}$ and nondipolar phase $\bm{\psi}_{N}^{\star}$
\end{algorithmic}
\end{algorithm}
A similar solver can be implemented using total generalized variation (TGV)\cite{bredies2013tgv}. The ADMM variables are chosen as $\mu_{1}=100\alpha$, $\mu_{2}=1$, $\mu_{\text{dip}}=10\lambda$.

The $\bm{\chi}$-update denominator $\mu_2 \mathbf{D}^{\mathsf{H}}\mathbf{D} + \mu_1 \mathbf{E}^{\mathsf{H}}\mathbf{E}$ vanishes at $\mathbf{k}=\mathbf{0}$, since both the dipole kernel and the finite-difference operator are zero there. A small numerical constant ($10^{-12}$) is added to this denominator to avoid division by zero, following the FANSI ADMM solver.

Because the Dipole-let filter bank is redundant and its scales overlap, reconstructing $\mathbf{z}_{\text{dip}}$ by summing the independently soft-thresholded per-scale coefficients is an approximate solution of the $\ell_1$-analysis proximal subproblem in Equation~\ref{eqn:dipoleletLagrangian}. It agrees with the standard scale-wise shrinkage-and-sum heuristic used throughout the undecimated wavelet denoising literature. It is exact only in the special case of non-overlapping (orthogonal) scale.

\newpage
\subsection{On choosing the Dipole-let Regularizer parameters}
The parameters $\alpha$, $\lambda$, and $\lambda_{\text{res}}$ act on different quantities, so their numerical values are not directly comparable. $\bm{\chi}$ and $\bm{\psi}_N$ compete to explain the same measured phase through $\mathbf{F}^{\mathsf{H}}\mathbf{D}\mathbf{F}\bm{\chi}+\bm{\psi}_N=\bm{\psi}$: any phase can, in principle, be attributed either to $\bm{\chi}$ (at a marginal cost governed by $\alpha$) or absorbed directly into $\bm{\psi}_N$ (at a marginal cost governed by $\lambda$, or $\lambda_{\text{res}}$ for its coarse component). If $\lambda$ is not sufficiently larger than $\alpha$, it becomes cheaper for the optimizer to route genuinely dipolar-consistent structure into $\bm{\psi}_N$ than to explain it through $\bm{\chi}$, which both erodes true susceptibility contrast and defeats the purpose of $\bm{\psi}_N$ as an estimate of the nondipolar component. Requiring $\lambda \gg \alpha$ keeps this competition tilted toward $\bm{\chi}$, so that $\bm{\psi}_N$ only absorbs phase content that is expensive to explain via the dipole model (genuinely nondipolar phase).

 Because $\lambda_{\text{res}}=15\alpha$ is numerically larger than $\lambda=10\alpha$, one might conclude that the heuristic regularizes the residual (coarse) component more strongly than the near-cone and outer scales. In practice is the opposite. First, in Algorithm~\ref{alg:dip} the ADMM variable $\mu_{\text{dip}}$ is itself set proportionally to $\lambda$: $\mu_{\text{dip}}=10\lambda$. The soft-thresholds applied are therefore $\lambda/\mu_{\text{dip}}=1/10$ for the $(j,m)$ scales and $\lambda_{\text{res}}/\mu_{\text{dip}}=(\lambda_{\text{res}}/\lambda)/10=0.15$ for the residual, a fixed $1.5\times$ ratio. Second, and more importantly, Dipole-let coefficients of a genuinely dipolar phase are not comparably scaled across scales to begin with, which is the precise mechanism the method relies on near-cone and outer coefficients are anomalously small for dipole-consistent phase, while the residual captures the bulk of the phase's smooth, low-frequency structure and is therefore substantially larger in magnitude  (main text, Section~2.4, ``Dipole-lets and Separating Streaking Artifacts''). The intended effect is that the heuristic regularizes substantially more aggressively near the magic cone, where nondipolar components dominates compared to true anatomy, than on the coarse residual, despite $\lambda_{\text{res}}>\lambda$ in  value.

\newpage
\subsection{Decomposition Filters}
\begin{figure}[htbp]
    \centering
    \includegraphics[width=1\textwidth]{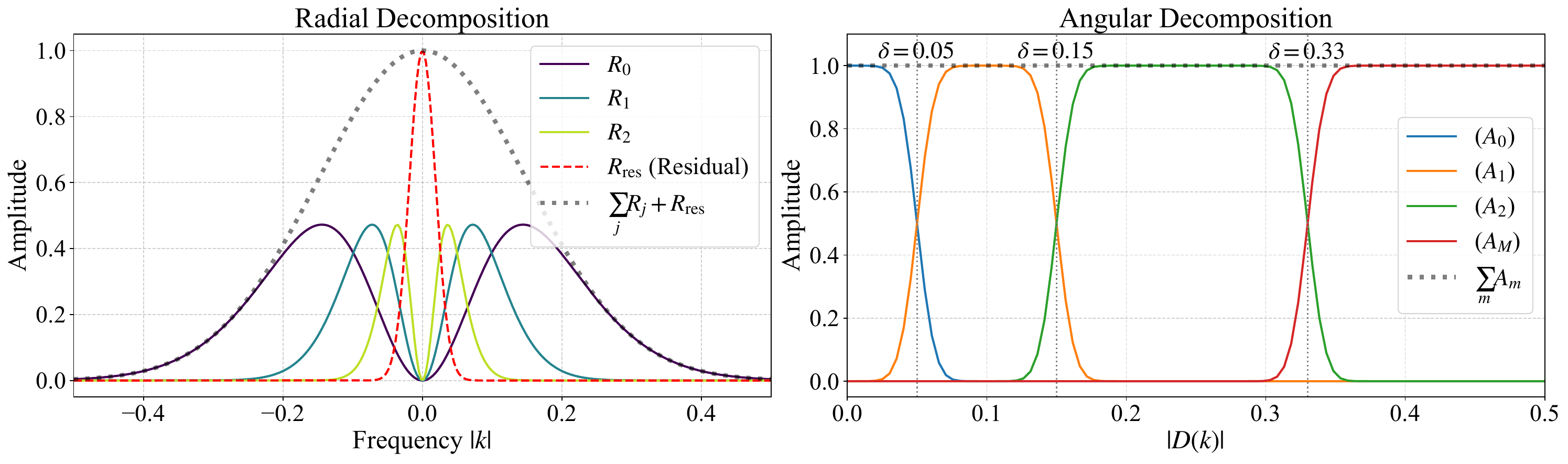}
     \caption*{\textbf{Figure S1.} Dipole-let filters for radial and angular decomposition in $1D$. (Left) Radial decomposition using Gaussian functions for $3$ scale and $1$ residual scale, $\sigma=0.15$, $\delta=0.05$ and $w=0.01$ . We also show the sum of all scales. (Right) Angular decomposition using the error function for 4 windows. The sum of all scales is equal to $1$ (partition of unity property).}
    \label{fig:dipoleletDecompositionNoisy}
\end{figure}
\subsection{Further Reconstruction Results}
\begin{figure}[htbp]
    \centering
    \begin{subfigure}[b]{0.33\textwidth}
        \centering
        \includegraphics[width=\textwidth]{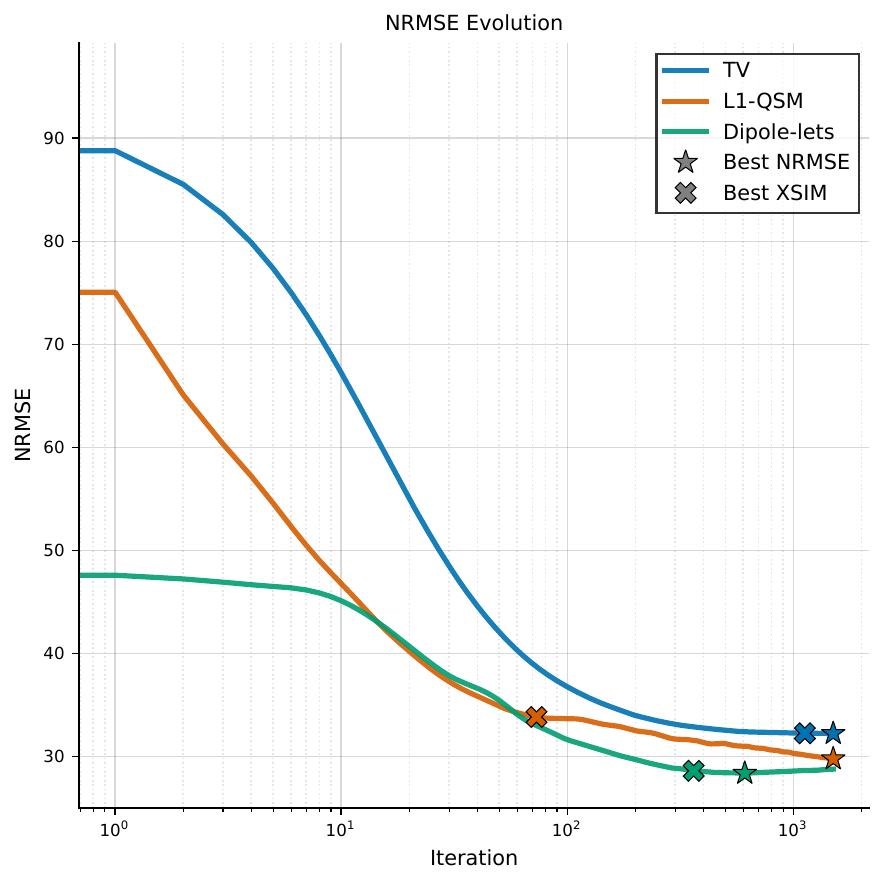}
        \caption{RMSE evolution.}
        \label{fig:sub1}
    \end{subfigure}
    \hfill
    \begin{subfigure}[b]{0.33\textwidth}
        \centering
        \includegraphics[width=\textwidth]{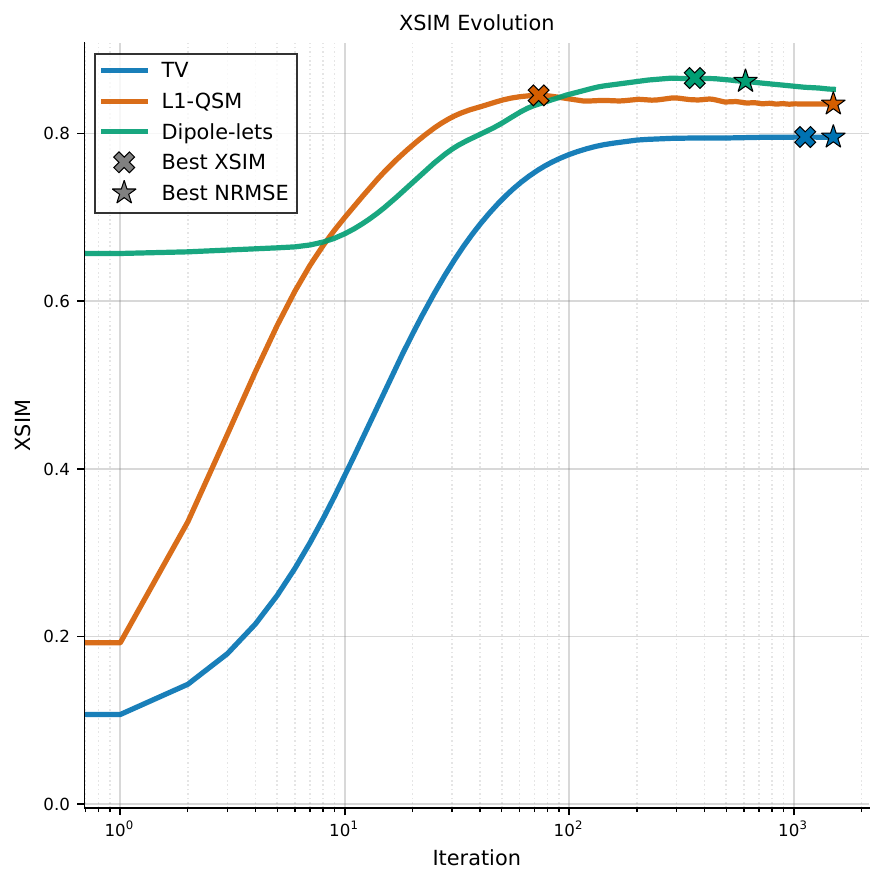}
        \caption{XSIM evolution.}
        \label{fig:sub2}
    \end{subfigure}
    \hfill 
    \begin{subfigure}[b]{0.33\textwidth}
        \centering
        \includegraphics[width=\textwidth]{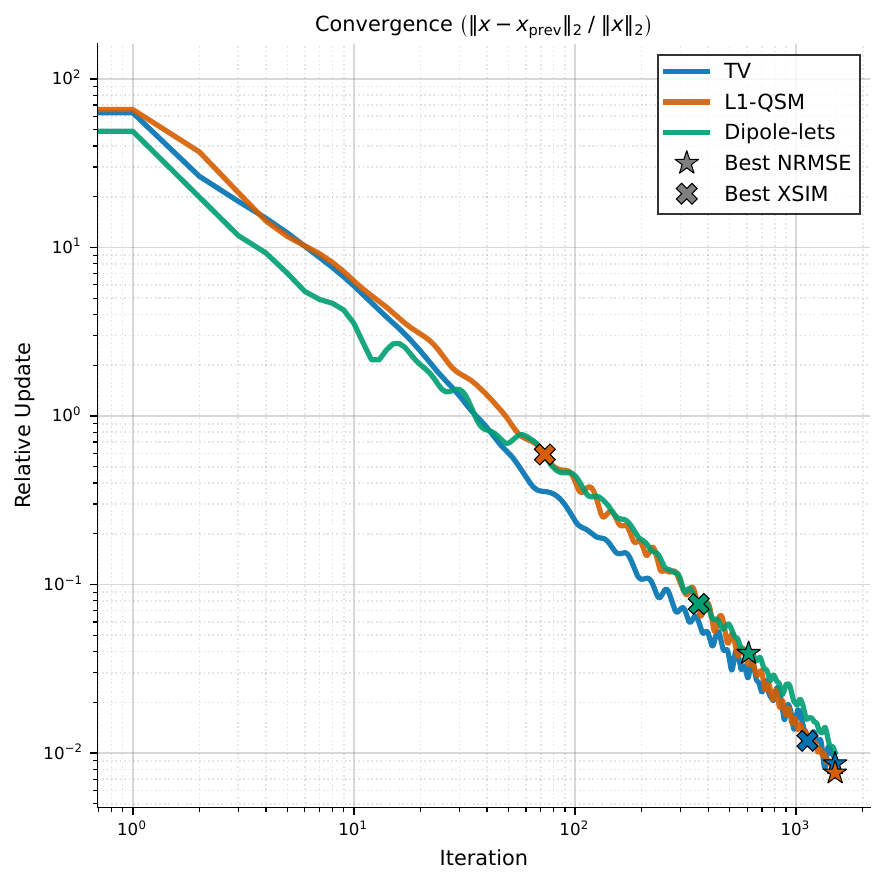}
        \caption{Convergence evolution.}
        \label{fig:sub3}
    \end{subfigure}
    
    \caption*{\textbf{Figure S2.} Evolution of RMSE, XSIM and convergence up to $1500$ iteration. Dipole-let converges to best RMSE early ($609$ iterations) compared to the other methods. After best RMSE and XSIM is achieved, both curves stays below and above the corresponding ones of L1-QSM and TV.}
    \label{fig:main_figure}
\end{figure}

\begin{figure}[htbp]
    \centering
    \includegraphics[width=0.7\textwidth]{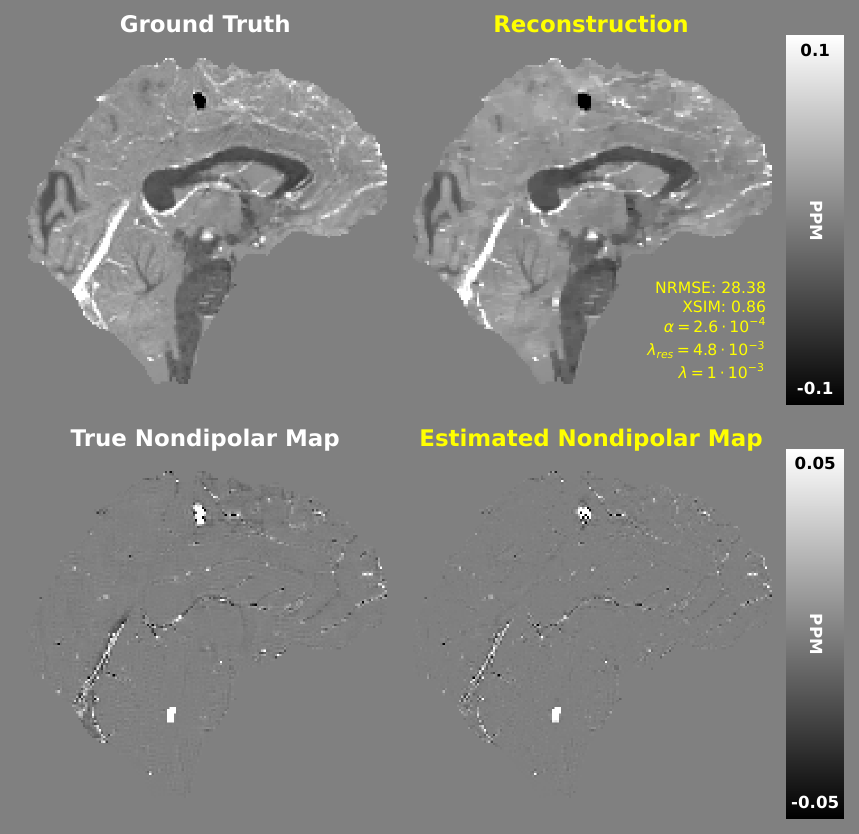}
     \caption*{\textbf{Figure S3.} Nondipolar phase component and its recovery by the Dipole-let Regularizer. Top row shows the ground truth (left) and the reconstruction (right). Bottom row shows the ground truth nondipolar phase component (left) and the nondipolar phase component $\bm{\psi}_N$ estimated by the Dipole-let Regularizer (right). Estimation shows high agreement, indicating the dipole-let transform correctly isolate the dipole-incompatible phase component from true dipolar phase. }

\end{figure}

\begin{figure}[htbp]
    \centering
    \includegraphics[width=1\textwidth]{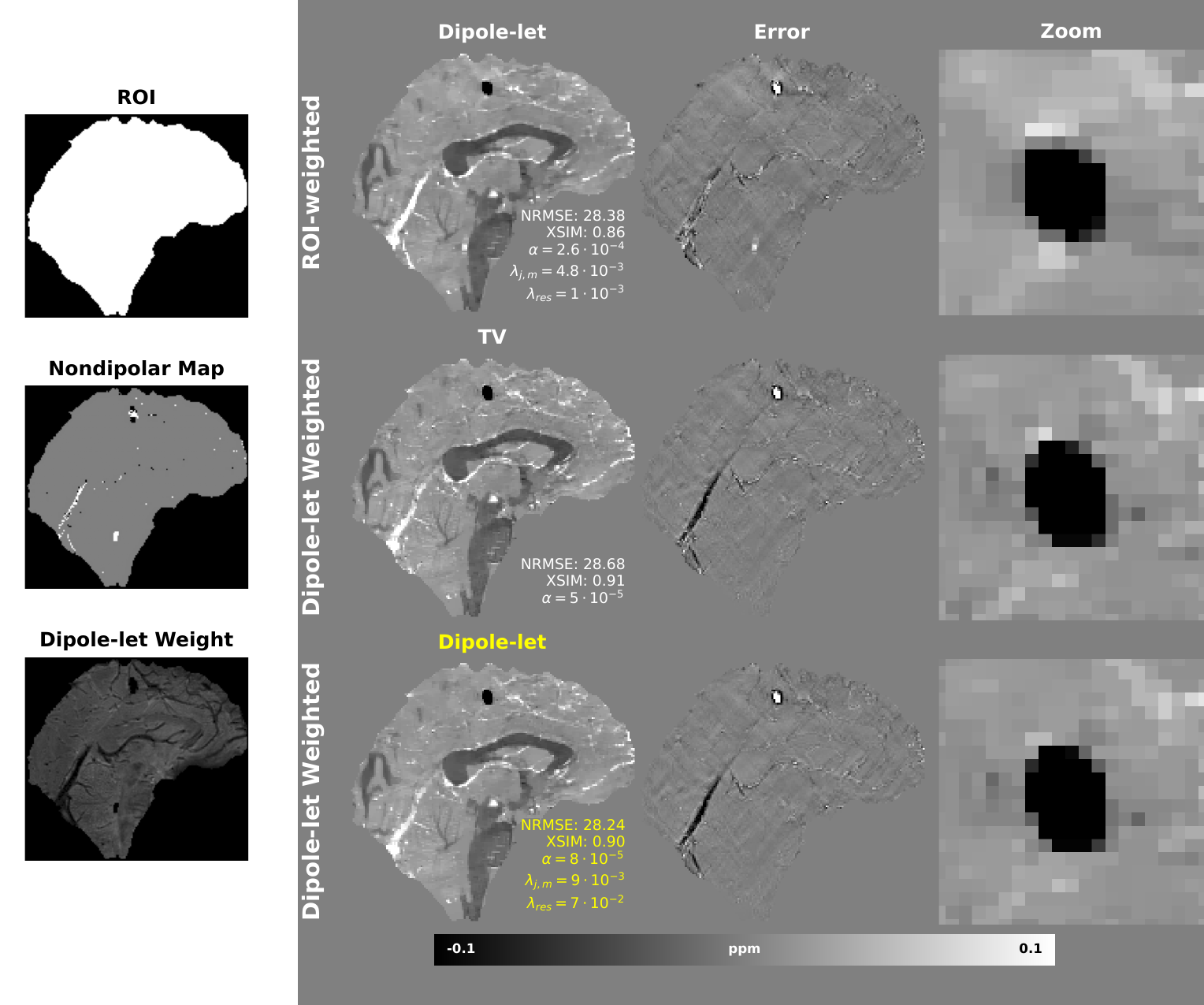}
     \caption*{\textbf{Figure S4.} Comparison of the Dipole-let weight and the Dipole-let regularizer, and the combination of both. First row shows Dipole-let regularizer (ROI-weighted) achieving second-best RMSE result but a zoom of the main calcification and added artifact shows still a minimal streaking. Dipole-weighting TV provides better XSIM results a better main calcification of streaking suppression. The combination of both, Dipole-let regularizer and Dipole-let weight give a slight improvement in RMSE and the XSIM improvement due to the suppression of the streaking of the main calcification and the added artifact. However the great vein (vein of Galen) is more attenuated compared to the ROI-weighted Dipole-let regularizer.}

\end{figure}

\begin{figure}[htbp]
    \centering
    \includegraphics[width=0.7\textwidth]{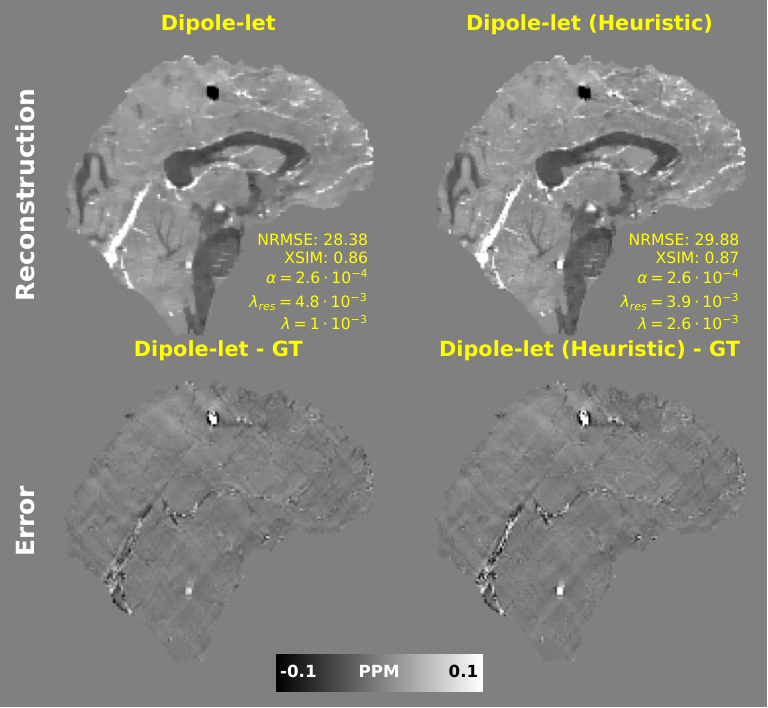}
     \caption*{\textbf{Figure S5.} Comparison of the Dipole-let Regularizer results using RMSE-optimal parameters versus the suggested heuristic ($\lambda = 10\alpha$ and $\lambda_{\text{res}} = 15\alpha$), and for $609$ iterations. As suggested, this heuristic provide comparable RMSE, XSIM and streaking reduction compared to best results.}
   
\end{figure}

\begin{figure}[htbp]
    \centering
    \includegraphics[width=0.9\textwidth]{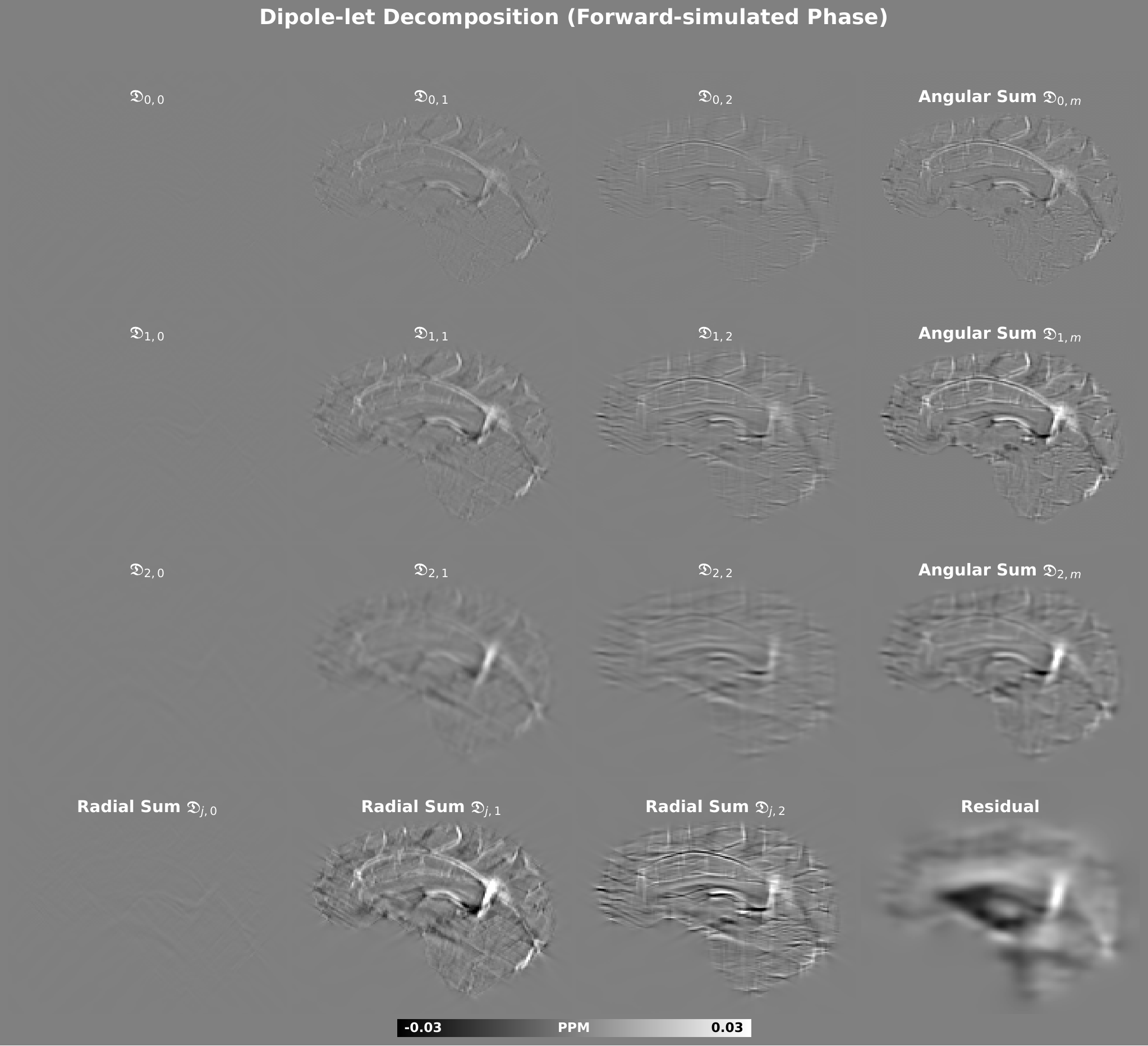}
     \caption*{\textbf{Figure S6.} Dipole-let decomposition ($J{=}3$ radial scales $\times$ $M{=}3$ angular bands, plus residual) of a forward-simulated, purely dipolar phase (RC1). At every radial scale, the near-cone column ($\mathfrak{D}_{j,0}$, leftmost) is visually flat, with no discernible anatomical structure, while the far-from-cone columns ($\mathfrak{D}_{j,1}$, $\mathfrak{D}_{j,2}$) and the residual carry essentially all of the phase's structure. Dipole-consistent phase has negligible near-cone Dipole-let energy. Compare directly with Figure~S7, the same anatomy after adding nondipolar perturbations.}

\end{figure}

\begin{figure}[htbp]
    \centering
    \includegraphics[width=0.9\textwidth]{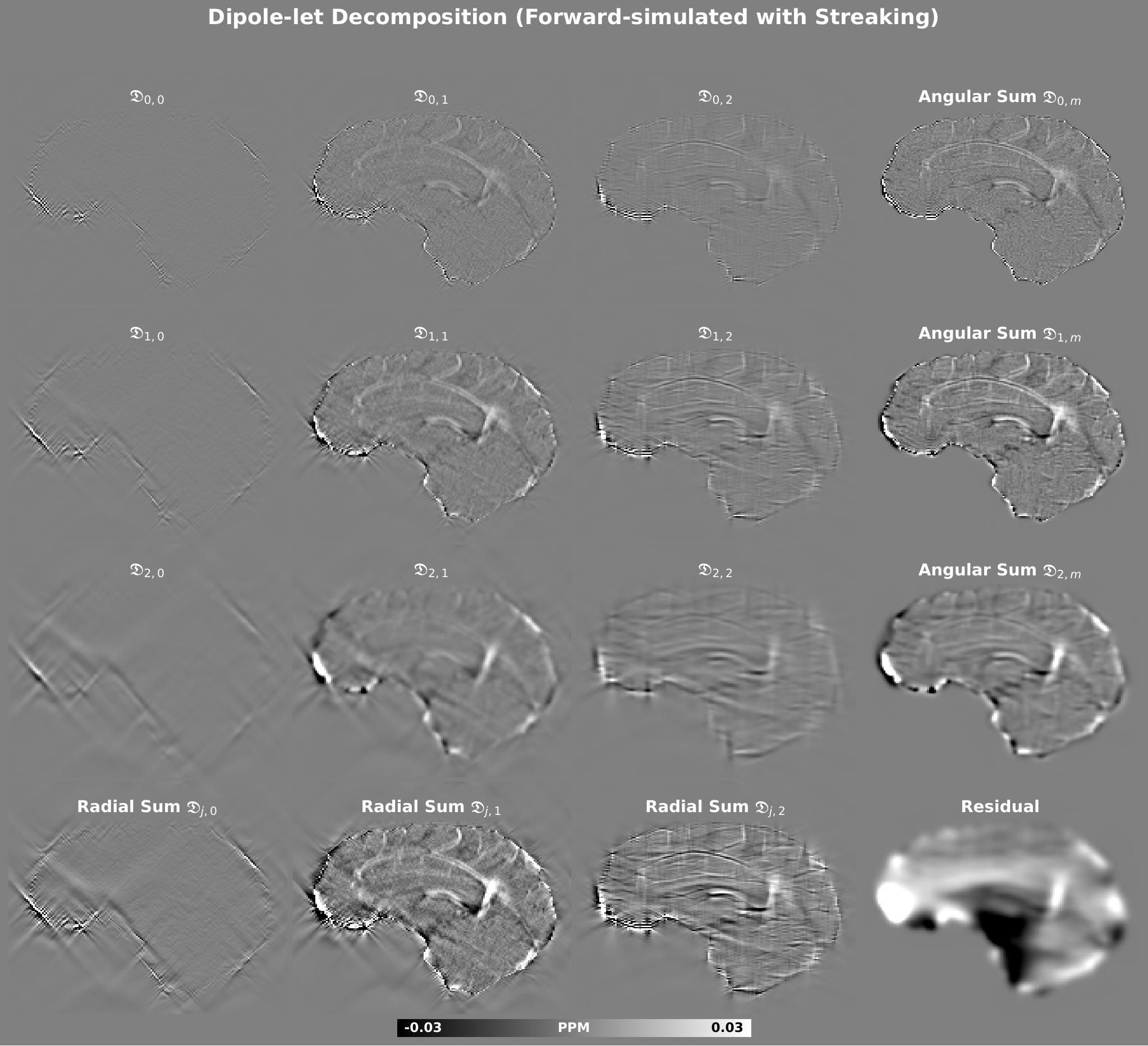}
     \caption*{\textbf{Figure S7.} Same decomposition and the same underlying anatomy as Figure~S6, after adding nondipolar perturbations to the forward-simulated signal: complex Gaussian noise (SNR $=100$) plus simulated background-field-removal remnants and boundary errors from masking. Unlike the clean case, the near-cone column ($\mathfrak{D}_{j,0}$) now shows high-energy response, concentrated along the brain boundary and following its contour at every radial scale, which is consistent with the mask (ROI) origin of the added perturbations.}

\end{figure}

\begin{figure}[htbp]
    \centering
    \includegraphics[width=0.8\textwidth]{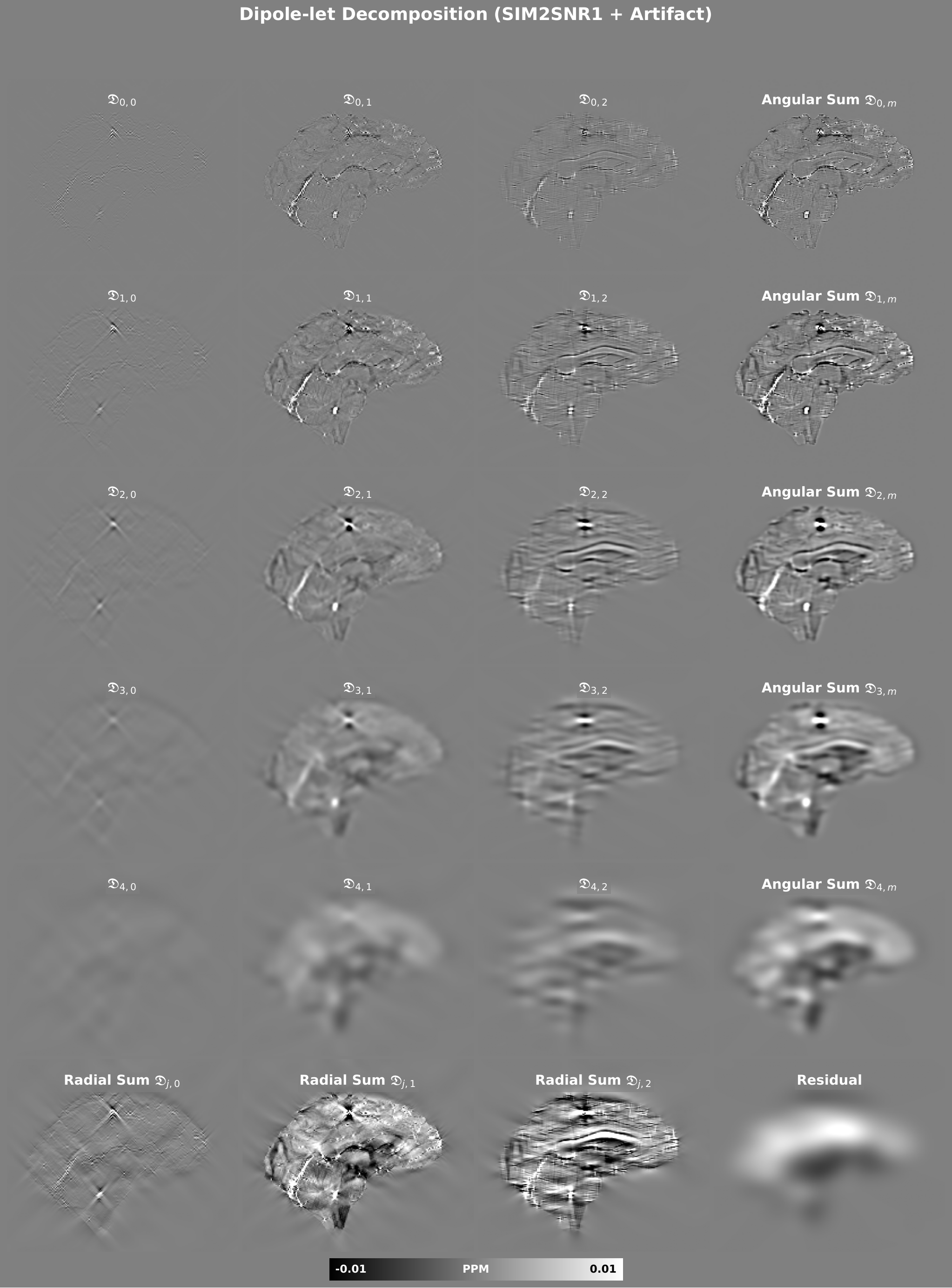}
     \caption*{\textbf{Figure S8.} Dipole-let decomposition ($J{=}5$ radial scales $\times$ $M{=}3$ angular bands, plus residual) of the modified SIM2SNR1 phase used throughout Sections~3.2--3.3 of the main text (calcification plus the added $1$\,Hz artifact near the pons). Two nondipolar phase components are visible in the near-cone bands: (i) an X-shaped pattern crossing the image center in $\mathfrak{D}_{2,0}$ and $\mathfrak{D}_{3,0}$ associated with the vein of Galen, and (ii) the small, sharply localized point responses near the top of the image (calcification) at several radial scales, and of the added artifact. The calcification response is consistent with it being a strong but spatially compact susceptibility source. Coefficients away from the cone ($\mathfrak{D}_{j,1}$, $\mathfrak{D}_{j,2}$) and the residual instead carry the expected smooth anatomical contrast. This is the same data and decomposition scales used to construct the fidelity weight shown in Figure~3 of the main text.}

\end{figure}

\begin{figure}[htbp]
    \centering
    \includegraphics[width=0.9\textwidth]{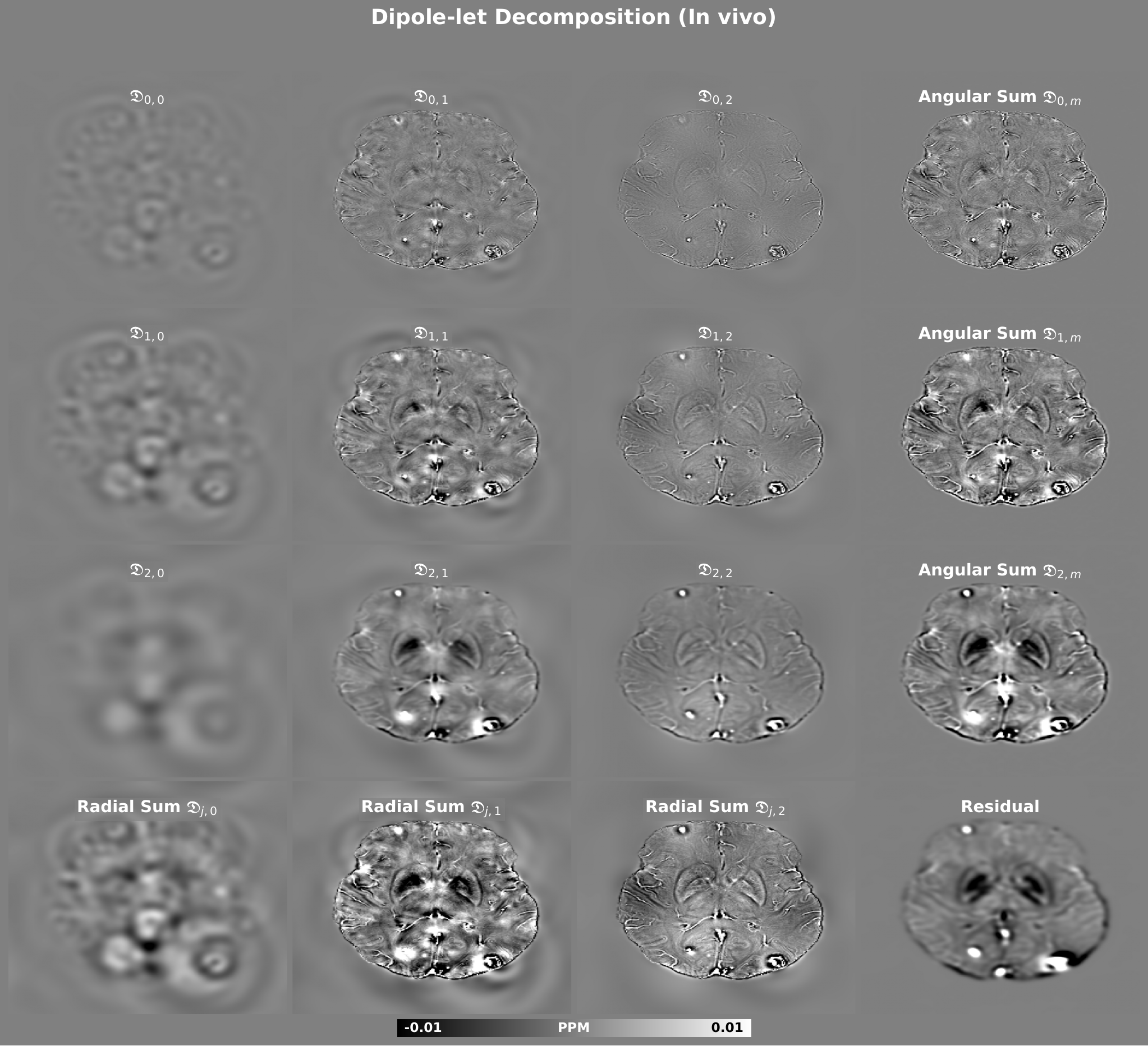}
     \caption*{\textbf{Figure S9.}  Dipole-let decomposition of the in vivo hemorrhage case (axial view, through the main hemorrhage). Unlike the simulated cases (Figures~S6--S8), the near-cone column ($\mathfrak{D}_{j,0}$) here is not flat: it shows clear ring-like structure spatially co-located with the hemorrhage. A large, strongly paramagnetic lesion such as a hemorrhage can itself produce genuine near-cone energy, through mechanisms (such as intravoxel dephasing or T2*-related effects) that are not processing artifacts but are also not captured by the scalar dipole model used to derive the method. This is consistent with the main text's discussion (Section~4) that dipole-lets can also flag genuine, clinically relevant tissue where the scalar dipole approximation itself is limited.}

\end{figure}

\end{document}